\documentstyle[aas2pp4]{article}
\newcommand{\ltwid}{\mathrel{\raise.3ex\hbox{$<$\kern-.75em\lower1ex\hbox{$\sim$
}}}}
\newcommand{\bd}{\begin{description}}
\newcommand{\ed}{\end{description}}
\newcommand{\s}{\scriptscriptstyle}
\begin{document}
\title{On Microlensing Event Rates and Optical Depth toward the
Galactic Center}
\author{S.J. Peale}
\affil{Dept. of Physics\\
University of California\\
Santa Barbara, CA 93106\\
peale@io.ucsb.edu}
\begin{abstract} 
The dependence of microlensing time scale frequency distributions and
optical depth toward the galactic center on galactic model parameters
is explored in detail for a distribution of stars consisting of the
Zhao (1996) bar and nucleus and the Bahcall and Soneira (1980) double
exponential disk.  The high sensitivity of these two microlensing
measures to the circular velocity model, velocity dispersions, bulge
mass, direction of the line of sight, bar axis orientation, star
spatial distribution 
and the stellar mass function means no single galaxy property can be
constrained very well without constraining most of the
others. However, this same sensitivity will make microlensing a
powerful member of the suite of observational techniques that will
eventually define the galaxy properties.  The model time scale
frequency distributions are compared with that determined empirically
by the MACHO group throughout. Although the MACHO
empirical data are matched quite well with a nominal velocity model and
with a mass function only of hydrogen burning stars that varies as 
$m^{-2.2}$ to $m^{-2.5}$ in the M star region, 
uncertainties in galactic structure, kinematics and content together
with the paucity of published microlensing data preclude any claim of
the model representing the real world. A variation of the mass
function $\sim m^{-1}$ in the M star region obtained from recent star
counts, both local and in the galactic bulge, fails to yield a
sufficient number of short time scale events compared to the MACHO
data.  The high sensitivity of the microlensing measures to the
direction of the line of sight may mean that sufficient microlensing
data is already in hand to constrain the bar distribution of stars.
The procedure developed here for determining the time scale frequency
distribution is particularly convenient for rapidly incorporating
model changes as data from all sources continues to accumulate. 
\end{abstract}
\section{Introduction}
The successful development of techniques to monitor microlensing
events toward the center of the galaxy has provided an additional
observational probe of galactic properties. Already an optical depth
that exceeds predictions (Alcock {\it et al} 1996; Udalski {\it et
al} 1994) has confirmed other indications (Dwek {\it et al} 1995)
that the galactic bulge must be bar shaped with the long axis pointing
toward us (Paczy\'nski {\it et al} 1994).  More massive lenses tend to
have longer time scales, so microlensing places  statistical
constraints on the mass function of the stellar lenses, although other
variables entering into the time scale weaken these
constraints. Attempts to define the mass 
function of the lenses from statistical arguments or from the event
time scale frequency distribution have led to models with a large
population of brown dwarfs in the bulge (Han and Gould 1996; Han
1998) but other models exist with few or no brown dwarfs (Zhao, Rich
\& Spergel 1996; Zhao and deZeeuw 1997; Mera, Cabrier \& Schaeffer
1998).  By
measuring optical depth and the distribution of event time scales as a
function of the line of sight direction, microlensing can also
constrain the distribution of stars in the bar. So far
there is insufficient data published [40 secure events from the MACHO 
(MAssive Compact Halo Objects) group (Alcock {\it et al} 1996) and 12
from the OGLE (Optical Gravitational Lensing Experiment)] group
(Udalski {\it et al} 1994)] for such constraints to be very
definitive. In fact, the lack of sufficient observations of all types
that can constrain the properties of the galaxy  means that few
of the potential constraints on galactic structure, kinematics or content
that microlensing can contribute to can be yet realized---the major
exception perhaps being the confirmation of the bar.

The microlensing optical depth is the probability that a  microlensing
event will be occurring at any instant in a particular direction for a
single source. The time scale frequency is the number of events of a
particular duration that occur per unit time, where the unit of time
can be the entire observational period and the number of sources can be
all of those monitored.  These two measures are routinely obtained
from a microlensing observational program aimed at monitoring stars
toward the galactic center, and form the basis for constraining
galactic properties.  It is important to appreciate the sensitivity 
of such data to the various galactic parameters on which they depend,
since they are related to the galaxy through models with assumed
values of the parameters.  Most of the galactic parameters such as
stellar velocities as a function of position are poorly
constrained observationally, which may allow several ways to
match microlensing data with very different galactic models. Given
this uncertainty, we wish to investigate the sensitivity of time scale
frequency distributions and optical depths to changes in the values of
several galactic parameters, and compare the model microlensing
measures at every step with the empirical measures deduced from the
microlensing events followed by the MACHO group during the summer of
1993 (Alcock {\it et al} 1996).  We eliminate the two stars that are
likely to be variable stars, two stars that might fail to pass the
MACHO criteria for inclusion and one
binary star from the data set. This leaves a rather meager data set of
only 40 events, but we chose not to add the 12 OGLE events in the
interest of having a data set all of whose entries are analyzed the
same way. The small gain in the statistics from adding the OGLE events
does not warrant the added complication.

The best
model of the distribution of stars in the galactic bulge is that due
to Zhao (1996). This model of the bar-shaped galactic bulge is based
on a model of Dwek {\it et al} (1994) that reproduces the COBE IR
surface brightness distributions, but which is also consistent with
all other available observations of bulge kinematics. A highly peaked
galactic nucleus is added to the triaxial ``boxy'' Gaussian
distribution of the bar to produce the observed central kinematics. 
We use the Zhao distribution of stellar mass (bar plus nucleus) and the
double exponential disk distribution of stars of Bahcall and Soneira
(1980) as a basis for determining microlensing event time scale
frequency distributions and microlensing optical depths as functions
of galactic parameters and direction of the line of sight.

In Section 2 we develop a general scheme for determining time scale
frequency distributions, in which it is relatively easy to substitute a
variety of stellar spatial distributions (although we limit these
particular changes to different bar orientations),  stellar velocities
that depend on position in the galaxy, mass functions, luminosity
functions and line-of-sight directions.  In Section 3 we discuss
galactic models in general and give details of the particular model
that will be the basis for the calculations. A short Section 4 describes
the  Monte Carlo integration scheme for two choices of independent
variables in the six dimensional integral defining the time scale
frequency distribution.  Section 5 is a series of example
calculations, where the first example is a demonstration of the
robustness of the time scale frequency distribution obtained with
either choice of independent variables. The remaining examples show
the effects of changing the visibility of the sources, the orientation
of the long axis of the galactic bar relative to the line from the
observer to the galactic center, the velocity model, the velocity
dispersions within one model, the mass function and the direction of
the line of sight.  The empirical time scale frequency distribution is
superposed on most of the graphs for comparison.  The last example in
Section 5 demonstrates an almost perfect match of the average of 
model time scale frequency distributions over the 24 MACHO field
directions with the empirical distribution.

A brief derivation of the optical depth averaged over the source
distribution along the line of sight begins Section 6.  This
expression is used to demonstrate a relatively large variation in
optical depth over a region that includes Baade's window.  The spatial
distribution 
of optical depths along each MACHO line of sight is compared with the
average optical depth determined empirically from the MACHO
observations, where the same model giving such a good match to the
empirical time scale frequency distribution gives an average optical
depth that is about $2\sigma$ below the empirical mean. The average 
of the model optical depths over the MACHO fields is raised almost to
the empirical mean if the bulge mass is multiplied by 1.5. The variation
of optical depth for a particular
direction on source visibility and orientation of the bulge axis is also
determined. In both Sections 5 and 6, we show the contributions of each
separate part of the distribution of stars, which demonstrates how
little the disk contributes to both the event rate and optical depth
relative to the bulge.  The discussion in Section 7 points out virtues
of and weaknesses in our particular formulation, emphasizes the
sensitivity of results to assumptions within the models, cautions the
reader about taking our model match to the data too seriously given
the large uncertainties in the real galactic model,
but points out the eventual power of microlensing to play a
major role in the entire suite of observations that will define the
properties of the galaxy in the future.  Our conclusions are
enumerated in a brief Section 8. 

\section{Event time scale frequency distribution}
A microlensing event is said to occur whenever a source star and lens
star pass each other at an angular separation within that of the
Einstein ring radius $R_{\s E}$ of the lens. The time scale for such
an event is defined as $t_{\s E}=R_{\s 
E}/v$, where $v$ is the magnitude of the relative transverse velocity
between source and lens projected onto the lens plane. This projected
transverse relative velocity is given by 
\begin{equation}
{\bf v}={\bf v}_{\s L}-{\bf v}_{\s O}-({\bf v}_{\s S}-{\bf v}_{\s 
O})\frac{D_{\s OL}}{D_{\s OS}}, \label{eq:vel}
\end{equation}
where $\bf v_{\s L},\,v_{\s S}$ and $\bf v_{\s O}$ are the transverse
velocities of the lens, source and observer respectively relative to
inertial space, and $D_{\s OL}$ and $D_{\s OS}$ are the distances to
the lens and source. The Einstein ring radius is 
\begin{equation}
R_{\s E}=\sqrt{\frac{4Gm}{c^2}\frac{D_{\s OL}(D_{\s OS}-D_{\s OL})}
{D_{\s OS}}}, \label{eq:re}
\end{equation}
where $G$ is the gravitational constant, $m$ is the lens mass and $c$ is
the velocity of light. 

A single source at distance $D_{\s OS}$ will lead to microlensing
events with lenses within mass range $dm$ of mass $m$, within $dD_{\s
OL}$ of $D_{\s OL}$, and with Einstein ring radii $R_{\s E}$ at the
rate of 
\begin{eqnarray}
2vR_{\s E}\frac{dn_{\s L}(m,D_{\s OL})}{dm}dD_{\s OL}dm&=&\cr
\frac{8v}{R_{\s E}}\frac{G}{c^2}
\frac{D_{\s OL}(D_{\s OS}-D_{\s OL})}{D_{\s
xOS}}
\frac{d\rho_{\s L}(m,D_{\s OL})}{dm}dD_{\s OL}dm,&& \label{eq:frate}
\end{eqnarray}  
where $(dn_{\s L}/dm)dm$ is the number density of lenses with masses within
$dm$ of $m$ so that $(dn_{\s L}/dm)dD_{\s OL}dm $ is the areal density of
lenses in the mass range $dm$ in the slab of thickness $dD_{\s OL}$ at
$D_{\s OL}$. The rate is expressed in terms of mass density 
of lenses through $d\rho_{\s L}(m,D_{\s OL})/dm=m\,dn_{\s L}(m,D_{\s
OL})/dm$, where the $m$ in the denominator cancels the $m$ in $R_{\s
E}^2$ in the second form. The transverse velocity $\bf v$ has
components $v_{\s b}$ and $v_{\s\ell}$ in the directions of increasing
galactic latitude and longitude respectively, such that 
\begin{equation} 
v=(v_{\s b}^2+v_{\s\ell}^2)^{1/2} \label{eq:tspeed}
\end{equation}
with
\begin{eqnarray}
v_{\s b}&=&v_{\s Lb}-v_{\s 0b}-(v_{\s Sb}-v_{\s 0b})D_{\s
OL}/D_{\s OS},\cr
v_{\s\ell}&=&v_{\s L\ell}-v_{\s 0\ell}-(v_{\s S\ell}-v_{\s 0\ell})D_{\s
OL}/D_{\s OS},\cr
v_{\s 0b}&=&v_{\s 0b}^{\s\prime},\cr
v_{\s 0\ell}&=&v_{\s LSR}+v_{\s 0\ell}^{\s\prime},\cr
v_{\s Lb}&=&v_{\s Lb}^{\s\prime},\cr
v_{\s L\ell}&=&v_{\s L0}+v_{\s L\ell}^{\s\prime},\cr
v_{\s Sb}&=&v_{\s Sb}^{\s\prime},\cr
v_{\s S\ell}&=&v_{\s S0}+v_{\s S\ell}^{\s\prime}.\label{eq:vdetails}
\end{eqnarray}
In Eqs. (\ref{eq:vdetails}), $v_{\s LSR}$ is the circular velocity of
the local standard of rest, $v_{\s 0b,\ell}$ are the peculiar velocity
components of
the observer relative to the LSR, $v_{\s L0}(D_{\s OL})$ is the
circular velocity of the galaxy at $D_{\s OL}$, $v_{\s
Lb,\ell}^{\s\prime}$ are the random lens velocities relative to the
mean galactic rotation, $v_{\s S0}(D_{\s OS})$ is the circular
velocity of the galaxy at $D_{\s OS}$, and $v_{\s 
Sb,\ell}^{\prime}$ are the random source velocities. We shall separate the
distribution of stars in the galaxy into disk and triaxial bulge
distributions with their own circular velocity distributions and
velocity dispersions, where the latter are Gaussian distributed about
the mean circular velocities. 

The number of source stars in a given field that are visible in range
$dD_{\s OS}$ at $D_{\s OS}$ is proportional to $n_{\s S}(D_{\s OS})D_{\s
OS}^{2+2\beta}$, where $n_{\s S}(D_{\s OS})$ is the number density of
sources at $D_{\s OS}$, and where the exponent follows from the
increase in area of the 
cross section of the angular field with distance and from the
assumption that the fraction of stars with luminosities greater than some
$L_*$ varies as $L_*^{\beta}$ (Kiraga and Paczy\'nski
1994). Hence, the fraction of the visible sources in the field that
are within the slab of thickness $dD_{\s OS}$ is given by 
\begin{displaymath}
n(D_{\s OS})D_{\s OS}^{2+2\beta}dD_{\s OS}{\mbox{\Huge
/}}\!\!\int_0^\infty n(D_{\s OS})D_{\s OS}^{2+2\beta}dD_{\s OS}
\end{displaymath} 

The expression (\ref{eq:frate}) is constrained to lenses within
$dD_{\s OL}$ at $D_{\s OL}$ and to lens masses within $dm$ of $m$. To
further constrain the event rate to particular ranges of lens and
source random velocities and to a range of source distances $dD_{\s
OS}$, we must multiply by the respective fractions of lenses and
sources contained within the small range of each parameter. Hence
\begin{eqnarray}
dF&=&
\frac{8}{t_{\s E}}\frac{G}{c^2}\frac{D_{\s OL}(D_{\s OS}-D_{\s OL})}{D_{\s
OS}}\frac{d\rho_{\s L}(m,D_{\s OL})}{dm}f_{\s v_{\s Lb}^{\s\prime}}
f_{\s v_{\s L\ell}^{\s\prime}}\times\cr
&&f_{\s v_{\s Sb}^{\s\prime}}
f_{\s v_{\s S\ell}^{\s\prime}}n(D_{\s OS})D_{\s OS}^{2+2\beta}
dv_{\s Lb}^{\s\prime}dv_{\s L\ell}^{\s\prime}
dv_{\s Sb}^{\s\prime}dv_{\s S\ell}^{\s\prime}\times\cr
&&dD_{\s OL}dD_{\s OS}dm{\mbox{\Huge /}}\!\!
\int_0^\infty n(D_{\s OS})D_{\s OS}^{2+2\beta}dD_{\s OS} \label{eq:srate}
\end{eqnarray}
is the rate of events for a single source within $dD_{\s OS}$ of $D_{\s
OS}$ that also has random velocities within $dv_{\s Sb}^{\s
\prime}dv_{\s S\ell}^{\s\prime}$ of $(v_{\s Sb}^{\s\prime},v_{\s
S\ell}^{\s\prime})$ for lenses within $dD_{\s OL}$ of $D_{\s
OL}$ with random velocities within $dv_{\s Lb}^{\s\prime}dv_{\s
L\ell}^{\s\prime}$ of $(v_{\s Lb}^{\s\prime},v_{\s L\ell}^{\s\prime})$
and with mass within $dm$ of $m$. The Gaussian velocity distributions are 
represented by 
\begin{displaymath}
f_{\s v_{\s Xy}^{\s\prime}}=\frac{1}{\sqrt{2\pi}\sigma_{\s
Xy}}\exp{\left(-\frac{{v_{\s Xy}^{\s\prime}}^2}{2\sigma_{\s Xy}^2}\right)},
\end{displaymath}
where $\sigma_{\s Xy}$ is the rms value of $v_{\s Xy}^{\s\prime}$
The circular velocities of the galaxy are assumed known from the
galactic model. The particular values of all seven
parameters in Eq. (\ref{eq:srate}) determine a particular value of the
time scale $t{\s E}=R_{\s E}/v$ which we have substituted.

To determine the total event rate per unit time scale interval for a
particular time scale $t_{\s E}=t_{\s E}^{\s\prime}$, we multiply
Eq. (\ref{eq:srate}) by $\delta(t_{\s E}-t_{\s E}^{\s\prime})$ and
integrate over the seven dimensional volume.  Before doing this, it is
expedient to normalize all distances by $D_{\s 8}=8$ kpc, the distance
to the galactic center, all velocity components by $v_{\s LSR}$, and
the masses by the solar mass $M_{\s\odot}$. The time scale $t_{\s E}$
is normalized by 
\begin{equation}
t_{\s 0}=\sqrt{\frac{4GM_{\s\odot}D_{\s 8}}{v_{\s
LSR}^2c^2}}=66.72\,{\rm days}
\label{eq:tnorm} 
\end{equation}
which is the time scale for an event with a solar mass lens located at
a distance of 4 kpc with the source at the galactic center (8
kpc). The source is assumed at rest and the lens velocity and observer
velocity are both assumed to be the circular velocity $v_{\s LSR}$ 
consistent with a flat galaxy rotation curve. The numerical value for
$t_{\s 0}$ is obtained with $v_{\s LSR}=210$ km/sec (Alvarez, May \&
Bronfmann  1990; Clemens 1985). These normalizations yield
\begin{eqnarray}
F&=&\frac{8\times 0^{33}}{(3.0856\times
10^{18})^3}\sqrt{\frac{GD_8^3v_{\s LSR}^2}{M_{\s\odot}c^2}}\int_{0.1}^{1.2} 
\int_0^\zeta\int_{0.08}^2\times\cr
&&\int_{-\infty}^\infty\int_{-\infty}^\infty\int_{-\infty}^\infty
\int_{-\infty}^\infty \frac{1}{t_{\s 
E}}\frac{z(\zeta-z)}{\zeta}\frac{d\rho_{\s L}(z,m)}{dm}\times\cr
&&f_{\s v_{\s L\ell}^{\s\prime}}f_{\s v_{\s Lb}^{\s\prime}}f_{\s v_{\s
S\ell}^{\s\prime}}f_{\s v_{\s Sb}^{\s\prime}} 
\zeta^{2+2\beta}n_{\s S}(\zeta)\delta(t_{\s E}-t_{\s
E}^{\s\prime})d\zeta\,dz\times\cr
&&dm\,dv_{\s L\ell}^{\s\prime}dv_{\s Lb}^{\s\prime}dv_{\s
S\ell}^{\s\prime}dv_{\s Sb}^{\s\prime}{\mbox{\Huge /}}
\!\!\int_{0.1}^{1.2}\zeta^{2+2\beta}n_{\s S}(\zeta)d\zeta \label{eq:ffrate}
\end{eqnarray}
as the number of events/sec/source/(unit $t_{\s E}$) at time scale
$t_{\s E}=t_{\s E}^{\s\prime}$. The numerical
coefficient means $\rho_{\s L}$ is expressed in $M_{\s\odot}/{\rm pc^3}$,
and $\zeta=D_{\s OS}/D_{\s 8}$, $z=D_{\s OL}/D_{\s 8}$, and it is
understood that all velocities are normalized by $v_{\s LSR}$ and
$t_{\s E}$ by $t_{\s 0}$. The integral limits correspond to the
variables in the same order as the differentials in the integrand. The
integral over $\zeta$ starts at 0.1 instead of 0 to avoid the
singularity in the integrand.  There are likely to be no sources
involved in microlensing events closer to us than 0.8 kpc. The upper
limit on $\zeta$ extends the range beyond the galactic center were
source visibility begins to fade from either distance or blending.
One can experiment with the effect of changing these limits, since
the extent of blending of stellar images depends on observing site
seeing. 

The integral is to be evaluated approximately with a Monte Carlo
technique, but first the dimension is reduced by one by changing
variables from $m,\,z,\,\zeta,\,v_{\s Lb}^{\s\prime},$ $v_{\s
L\ell}^{\s\prime},\,v_{\s Sb}^{\s\prime},\,v_{\s S\ell}^{\s\prime}$ to
$t_{\s E},\,z,\,\zeta,\,v_{\s Lb}^{\s\prime},\,v_{\s
L\ell}^{\s\prime},\,v_{\s Sb}^{\s\prime},\,v_{\s S\ell}^{\s\prime}$
and integrating over the $\delta$ function.  The Jacobian determinant
of the transformation is simply $\partial m/\partial t_{\s E}$. In
dimensionless units, 
\begin{equation}
t_{\s E}=\sqrt{\frac{mz(\zeta-z)}{\zeta(v_{\s b}^2+v_{\s\ell}^2)}},
\label{eq:te}  
\end{equation}
such that
\begin{equation}
\frac{\partial m}{\partial t_{\s E}}=\frac{2t_{\s E}(v_{\s
b}^2+v_{\s\ell}^2)\zeta}{z(\zeta-z)}
\end{equation}
The change of variables in the integrand
of Eq. (\ref{eq:ffrate}) is effected by replacing $dm$ with $(\partial
m/\partial t_{\s E})dt_{\s E}$. Integration over $t_{\s E}$ fixes 
$t_{\s E}$ at the value selected by the $\delta$ function, and the
remaining six dimensional integral is
\begin{eqnarray}
F&=&\frac{16\times10^{33}}{(3.0856\times
10^{18})^3}\sqrt{\frac{GD_8^3v_{\s LSR}^2}{M_{\s\odot}c^2}}\int_{0.1}^{1.2} 
\int_0^\zeta\int_{-\infty}^\infty\times\cr
&& \int_{-\infty}^\infty \int_{-\infty}^\infty
\int_{-\infty}^\infty(v_{\s b}^2+v_{\s\ell}^2)\frac{d\rho_{\s L}(z,m)}{dm}
f_{\s v_{\s L\ell}^{\s\prime}}f_{\s v_{\s
Lb}^{\s\prime}}f_{\s v_{\s S\ell}^{\s\prime}}\times\cr
&&f_{\s v_{\s Sb}^{\s\prime}}\zeta^{2+2\beta}n_{\s S}(\zeta)
d\zeta\,dz\,dv_{\s L\ell}^{\s\prime}dv_{\s Lb}^{\s\prime}dv_{\s
S\ell}^{\s\prime}dv_{\s Sb}^{\s\prime}{\mbox{\Huge /}}\cr
&&\int_{0.1}^{1.2}\zeta^{2+2\beta}n_{\s S}(\zeta)d\zeta, \label{eq:mcarlo}
\end{eqnarray}
where $t_{\s E}$ appears only in the expression for $m$ in $d\rho_{\s
L} (z,m)/dm$. 

As a check on the procedure, we eliminate $z$ instead of $m$ as an
independent variable in Eq. (\ref{eq:ffrate}). {\it I.e.},
$m,\,z,\,\zeta,\,v_{\s Lb}^{\s\prime},\,v_{\s 
L\ell}^{\s\prime},\,v_{\s Sb}^{\s\prime},\,v_{\s S\ell}^{\s\prime}$ go
to $m,t_{\s E},\,\zeta,\,v_{\s Lb}^{\s\prime},\,v_{\s
L\ell}^{\s\prime},\,v_{\s Sb}^{\s\prime},$ $v_{\s S\ell}^{\s\prime}$)
with $\partial z/\partial t_{\s E}$ being the only term in the
Jacobian determinant.
Implicit differentiation of the square of Eq. (\ref{eq:te}) yields  
\begin{equation}
\frac{\partial z}{\partial t_{\s E}}=\left[\frac{t_{\s E}(\zeta-2z)}
{2z(\zeta-z)}-\frac{t_{\s E}^3\zeta}{mz(\zeta-z)}\left(v_{\s b}\frac
{\partial v_{\s b}}{\partial z}+v_{\s\ell}\frac{\partial v_{\s\ell}}
{\partial z}\right)\right]^{-1} \label{eq:dzdte}
\end{equation}
where
\begin{eqnarray*}
\frac{\partial v_{\s b}}{\partial z}&=&(v_{\s 0b}^{\s\prime}-v_{\s
Sb}^{\s\prime})/\zeta, \cr
\frac{\partial v_{\s\ell}}{\partial z}&=&(-1+v_{\s 0\ell}-v_{\s
S\ell}^{\s\prime})/\zeta.
\end{eqnarray*} 
Some important assumptions are used in arriving at the above
forms of these partial derivatives that will be pointed out below.
From Eq. (\ref{eq:te}), we can write $az^2+bz+c=0$, with
\begin{eqnarray*}
a&=&m+t_{\s E}^2[(1-v_{\s S\ell})^2+v_{\s Sb}^2]/\zeta,\cr
b&=&-m\zeta+2t_{\s E}^2[v_{\s L\ell}(1-v_{\s S\ell})-v_{\s Lb}v_{\s
Sb}],\cr
c&=&t_{\s E}^2\zeta(v_{\s L\ell}^2+v_{\s Lb}^2),
\end{eqnarray*}
where $v_{\s Xy}=v_{\s Xy}^{\s\prime}-v_{\s 0y}$ are transverse
velocities relative to the observer. The solutions of the quadratic
equation for $z$ allow its expression in terms of the remaining
variables where it occurs in $\partial z/\partial t_{\s E}$ and
elsewhere in the integrand.  The change of variables in the integrand
of Eq. (\ref{eq:ffrate}) is effected by replacing $dz$ with $(\partial
z/\partial t_{\s E})dt_{\s E}$. Integration over $t_{\s E}$ fixes 
$t_{\s E}$ at the value selected by the $\delta$ function, and the
remaining six dimensional integral is
\begin{eqnarray}
F&=&\frac{8\times10^{33}}{(3.0856\times
10^{18})^3}\sqrt{\frac{GD_8^3v_{\s LSR}^2}{M_{\s\odot}c^2}}\int_{0.1}^{1.2} 
\int_{0.08}^2\int_{-\infty}^\infty\times\cr
&& \int_{-\infty}^\infty
\int_{-\infty}^\infty \int_{-\infty}^\infty \frac{1}{t_{\s
E}}\frac{z(\zeta-z)}{\zeta}\times \cr
&&\frac{d\rho_{\s L}(z,m)}{dm}
f_{\s v_{\s L\ell}^{\s\prime}}f_{\s v_{\s
Lb}^{\s\prime}}f_{\s v_{\s S\ell}^{\s\prime}}f_{\s v_{\s Sb}^{\s\prime}}
\zeta^{2+2\beta}n_{\s S}(\zeta)\times\cr
&&\frac{\partial z}{\partial t_{\s
E}}(t_{\s_E},\zeta,v_{\s Lb}^{\s\prime},v_{\s L\ell}^{\s\prime},v_{\s 
Sb}^{\s\prime},v_{\s S\ell}^{\s\prime})\times \cr
&&d\zeta\,dm\,dv_{\s L\ell}^{\s\prime}dv_{\s Lb}^{\s\prime}dv_{\s
S\ell}^{\s\prime}dv_{\s Sb}^{\s\prime}{\mbox{\Huge /}}\cr
&&\int_{0.1}^{1.2}\zeta^{2+2\beta}n_{\s S}(\zeta)d\zeta \label{eq:mcarlo2}
\end{eqnarray}
With fixed $t_{\s E}$, the solutions of the quadratic equation for
$z$, the fractional distance of the lens to the center of the galaxy,
for random choices of the independent variables within their physical
ranges must be real with the added condition that $z<\zeta$.  Those
sets of variables for which there are no real solutions for $z$
are simply incompatible with the particular
value of $t_{\s E}$.  The two expressions for $F$,
Eqs. (\ref{eq:mcarlo}) and (\ref{eq:mcarlo2}), should yield identical
time scale frequency distributions.  We shall demonstrate this below,
but use the simpler form, Eq. (\ref{eq:mcarlo}) in subsequent
calculations, wherein it is easy to introduce additional $z$ and
$\zeta$ dependence into the velocities.  
\section{Models}
There is an arbitrary degree of complexity that one may
incorporate into the galactic models that specify the distributions of
stars and their velocities, and many of the parameters are not well
constrained by observations. The random velocities $v_{\s
Lb,\ell}^{\s\prime}=v_{\s Lb,\ell}^{\s\prime}(m,z,b,\ell)$ in general
with similar dependencies on parameters for $v_{\s Sb,\ell}^{\s\prime}$.
The dependency of the velocity dispersion on stellar mass $m$ is
consistent with the increasing thickness of the spatial distribution
in the disk with decreasing mass ({\it e.g.} Mihalas and Binney,
1981). The  best model of the galactic bar consistent with the COBE
IR intensity distribution (Zhao 1996) shows explicitly the dependence
of the dispersions on $b$ and $\ell$. Part of the dependence of the
velocities on $z$ can be accounted for by separating the galaxy into
disk and bulge regions with different velocity dispersions in each
region but with the dispersions being otherwise independent of $z$ and
$\zeta$ respectively for lenses and sources.  This latter assumption
was used explicitly in evaluating $\partial v_{\s b,\ell}/\partial z$ in
Eq. (\ref{eq:dzdte}), where the $z$ dependence of the velocities was
ignored. We will relax this assumption in some of the examples below.

The mean circular velocity of the gas clouds in the galactic disk 
appears to be a function of $z$ or $\zeta$ (Clemens 1985), but the
rather complex rotation curve is usually approximated as flat with all
mean circular velocities equal to $v_{\s LSR}$ (Alvarez {\it et al} 1990).
(Here we ignore any peculiar velocity of the $LSR$ relative to the
local mean circular velocity (Clemens 1985).)
The mean circular velocity of the stars is assumed to be the same as
that of the gas clouds.  A distribution of box-like orbits in the Zhao
(1996) bar model matches the observed HI velocity distribution, and
predicts the velocity dispersions that should be expected as a
function of $b$ and $\ell$ and presumably as a function of $z$ or
$\zeta$ inside the bulge.      

Microlensing should eventually prove a powerful constraint on the
stellar mass function for the stellar spatial distribution that
dominates the lens population, which is apparently the central bar.
This is especially so for the lower mass stars, since there is
no other way of observing these stars. Generally, one expects the mass
function to depend on $z$ (or $\zeta$) in general, but that dependence
is unknown. The frequency of event time scales is the microlensing
constraint on the mass function, where the fraction of shorter time
scales increases as the fraction of the mass density that is in the
smallest stars or brown dwarfs is increased.  However, increasing the
velocity dispersions or otherwise increasing relative transverse velocities
between lenses and sources and concentrating more of the lenses closer
to the center of the galaxy both increase the fraction of short time
scale events.  So constraints on dispersions and spatial distributions
of lenses and sources must be secure before definitive mass functions
can be ascertained. 

We have adopted above the assumption of Kiraga and Paczy\'nski (1994)
that the luminosity function of the sources is such that the fraction
of stars with luminosities greater than some luminosity $L_*$ is
proportional to $L_*^{\beta}$. This appears to be a fairly good
approximation over the important range of luminosities in K band
(Tiede, Frogel \& Terndrup 1995), but less so in the visual band where the
microlensing observations are made. More complex luminosity functions
may affect conclusions based on microlensing data but are not
warranted by observational constraints at this time. Data provided in
Tiede {\it et al} (1995) yield $\beta\approx -1$, which value is
assumed in most of the examples herein.

In principle, one could incorporate any of these complexities into the
galactic model, but there is little motivation for doing very much of
this until observational constraints become more secure. Even without
this security, one can determine the effect of various assumptions on
the microlensing time scale rate and optical depth (Zhao and deZeeuw
1998) in anticipation of future selection. For our purposes here, we
shall adopt a galactic model for the integration of
Eq. (\ref{eq:mcarlo}) that consists of a double exponential
disk and a bar-shaped bulge. The disk model is that of Bahcall and
Soneira, (1980).
\begin{equation}
\rho_{\s m}=\int_{m_{\s min}}^{m_{\s max}}
\frac{d\rho}{dm}dm=\rho_{\s
m0}\exp{\left(\frac{-|z^{\s\prime}|}{300{\rm pc}}-\frac{r}{s_{\s
d}}\right)},   \label{eq:disk}
\end{equation}
where $z^{\s\prime}$ is the coordinate perpendicular to the plane of
the galaxy, r is the radial coordinate in the plane of the galaxy, and
$s_{\s d}$ is the scale length in the radial direction, which may be
less than the 3.5 kpc used by Bahcall and Soneira (Zhao, Spergel \&
Rich 1995; Kent, Dame \& Fasio 1995), and where $\rho_{\s m0}$ is
chosen such that $\rho_{\s m}=0.05M_{\s\odot}/{\rm pc^3}$ at
$z^{\s\prime}=0$ and $r=8$ kpc.  Like Zhao {\it et al} (1996), we choose
the triaxial bulge model of Zhao (1996) but keep the nucleus and do
not terminate the bulge at 3.3 kpc.
\begin{equation}
\rho_m=\rho_{\s 0}\left[\exp\left(-\frac{s_{\s b}^2}{2}\right)+s_{\s
a}^{-1.85}\exp(-s_{\s a})\right], \label{eq:bulge}
\end{equation}
where 
\begin{eqnarray}
s_{\s b}^4&=&\left[\left(\frac{x}{\sigma_{\s x}}\right)^2+\left(\frac{y}
{\sigma_{\s y}}\right)^2\right]^2+\left(\frac{z^{\s\prime}}{\sigma_{\s
z}}\right)^4,\cr
s_{\s a}^2&=&\frac{q_{\s a}^2(x^2+y^2)+z^{{\s\prime}2}}{\sigma_{\s z}^2}
\label{eq:blgexp} 
\end{eqnarray}
with $q_{\s a}=0.6$, $\sigma_{\s x}=1.49$ kpc, $\sigma_{\s y}=0.58$ kpc and
$\sigma_{\s z}=0.40$ kpc. The coefficient $\rho_{\s 0}$ determines the
mass of the bulge.  The observations are consistent with the long axis
of the bulge inclined about $13^\circ$ to $20^\circ$ relative to the
line of sight to the galactic center with the near side of the bar
lying in the first quadrant.  There are large uncertainties in this
inclination angle, however (Zhao 1998). 

In the applications, the origin of
coordinates for both disk and bulge distributions will be
transferred to the position of the observer with the following
transformation.
\begin{eqnarray}
x&=&\cos{\theta}-\zeta\cos{b}\cos{(\ell-\theta)}\cr
y&=&-\sin{\theta}-\zeta\cos{b}\sin{(\ell-\theta)}\cr
z^{\s\prime}&=&\zeta\sin{b}  \label{eq:blgtrans}
\end{eqnarray}
for a source in the bulge with $xyz^{\s\prime}$ being galacto-centric
coordinates  along the bulge principal axes and $\theta$ is the inclination
of the bulge axis to the line between the observer and the galactic
center  measured in the direction of
increasing $\ell$.  If a bulge star acts as a lens, $\zeta$ is
replaced by $z$ in   Eqs. (\ref{eq:blgtrans}). For the disk
distribution, the transformation is
\begin{eqnarray}
r^2&=&1+z^2\cos^2{b}-2z\cos{b}\cos{\ell}\cr
z^{\s\prime}&=&z\sin{b} \label{eq:disktrans}       
\end{eqnarray}
Since $b$ and $\ell$ are always small angles, $\cos{b}\approx\cos{\ell}
\approx 1$ is usually adequate approximation, and the exponent in
Eq. (\ref{eq:disk}) can be replaced by $(1-s_{\s d}D_{\s 8}\sin{|b|}/300
{\rm pc})$ $D_{\s 8}z/s_{\s d}$. 

In some of the examples below, we shall assume the circular velocity
of the disk to be 210 km/sec 
independent of radial position $r$, with the random velocities
superposed on this base. The bar can be assumed uniformly rotating at
an angular velocity of $63.6 {\rm km/sec/kpc}$, which is slightly higher
than the $60 {\rm km/sec/kpc}$ pattern speed  chosen by Zhao (1996)
but is arbitrarily  chosen to match the disk rotation at 3.3 kpc. A
bulge velocity dispersion will be superposed on this bar
rotation. We shall also consider the case where the bar
is not rotating, but neither of these assumptions 
is consistent with the results of Alvarez {\it et al} (1990)
who find the flat rotation curve of the galaxy gas clouds (210 km/sec)
to persist down to 2 kpc from the center with velocities exceeding 210
km/sec at smaller galacto-centric distances.  However, there are
non circular streaming motions in the bar that could boost the
apparent rotation inside 3.3 kpc (Zhao 1996).  Finally, in most of our
examples we shall
consider  the case where  both circular velocities and velocity
dispersions for lenses 
and sources are functions of where they are located in the
galaxy. {\it E.g.}, Circular velocity $=210\,{\rm km/sec}$ and velocity
dispersions are those appropriate to the disk for stars further than
3.3 kpc from the galactic center, but would be those appropriate to
the bulge for smaller radial distances.

Because there is no compelling argument specifying the mass function
variation with galactic radius, we shall assume a single mass function
applies throughout the galaxy.  Most of the mass functions will be
limited to the range $0.08\leq m \leq 2.0M_{\sun}$. The upper limit is
chosen because there are so few stars with larger mass, and the lower
limit is chosen arbitrarily at the hydrogen burning limit.
The consequences of using several different
mass functions, sometimes extended into the brown dwarf region,
will be demonstrated in examples that illustrate the sensitivity of
the microlensing technique to the mass function's  eventual constraint. 

\section{Monte Carlo integration}
For the choice of variables $z,\zeta,v_{\s Lb},v_{\s L\ell},v_{\s
Sb},v_{\s S\ell}$ (Eq. (\ref{eq:mcarlo})) with no explicit $z$ or
$\zeta$ dependence for the dispersions,
the volume over which the integration variables are distributed randomly
in the Monte Carlo integration is chosen to be $(1.1)(1.2)(6^4)$
$\sigma_{\s Lb}\sigma_{\s L\ell}\sigma_{\s Sb}\sigma_{\s S\ell}$ so
that the random velocities are truncated at $\pm 3\sigma_{\s
Xy}$. For the examples where the dispersions and circular velocities
are functions of $z$ or $\zeta$, all of the $\sigma$'s in the expression
for the volume are replaced by the maximum among them.  Those random
values of $z>\zeta$ are outside the volume over 
which the integration is taken.  For the variables $m,\zeta,v_{\s Lb},
v_{\s L\ell},v_{\s Sb},v_{\s S\ell}$ (Eq. (\ref{eq:mcarlo2})), the
volume is $(1.92)(1.1)(6^4)\sigma_{\s Lb}\sigma_{\s L\ell}\sigma_{\s
Sb}\sigma_{\s S\ell}$. We choose $0.1\le\zeta\le 1.2$ since there are
not likely 
to be sources closer than 0.8 kpc and faintness and blending will
likely make sources beyond about 10 kpc unusable with the possible
exception of the clump giants.  The values of the six variables in the
integrand are chosen randomly for each evaluation of the integrand,
and values of $m$ in the first set and 
$z$  in the second as  functions of
these variables and the fixed $t_{\s E}$ are determined by the solutions of
Eq. (\ref{eq:te}). All points for which $m$ is not within the
specified range in the first form of $F$ or where there are no real
solutions for $z$ (the particular choice of random variables not being
consistent with the particular time scale) in the second or for
$z\ge\zeta$ do not contribute to the integral. In the second form, the
integral is evaluated separately for each of two viable values of $z$
and the results added to get the total contribution to 
the integral for a particular set of random variables.  A sufficient
number of points in the 6 dimensional volume are used in evaluating
the Monte Carlo integral (typically $10^6$) to yield a relatively
smooth frequency distribution as a function of $t_{\s E}$.  

\section{Examples}
\begin{figure}[ht]
\plotone{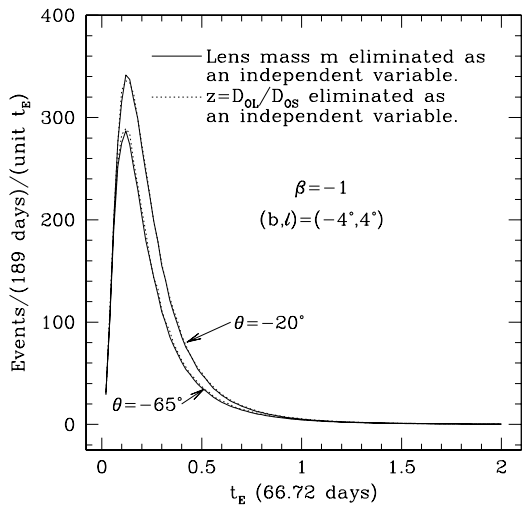}
\caption{Comparison of two Monte Carlo integrations with different
variables eliminated from the integrand. Zhao (1996) bar model with
nucleus ($M=2.2\times 10^{10}M_\odot$). Double exponential disk model
with scale lengths of 2.7 kpc and 300 pc and local mass density of
$0.05M_\odot/{\rm pc^3}$. $v_{\s L0}=v_{\s LSR}=210$ km/sec, $v_{\s
S0}=0$. $(\sigma_{\s Lb},\sigma_{\s L\ell},\sigma_{\s Sb},\sigma_{\s
S\ell})=(16,\,20,\,124,\,102)$ km/sec.  The ordinate corresponds to
12.6 million source stars monitored over 189 days. \label{fig:mvszcmpre}}   
\end{figure}

First we show that  both means of calculating the time scale
frequency distribution (Eqs. (\ref{eq:mcarlo}) and (\ref{eq:mcarlo2}))
yield the same result.  For this purpose, we assume $v_{\s L0}=v_{\s
LSR}=210$ km/sec, $v_{\s S0}=0$, $(\sigma_{\s Lb},\sigma_{\s
L\ell},\sigma_{\s Sb},\sigma_{\s S\ell})=(16,\,20,\,124,\,102)$ km/sec, 
$(b,\ell)=(-4^\circ,4^\circ)$, where the bulge velocity dispersions
correspond to those given for these coordinates by Zhao (1996), bulge
mass $=2.2\times 
10^{10}M_{\s\odot}$, local mass density in the disk model
$=0.05M_{\s\odot}/{\rm pc^3}$ with the Basu and Rana mass function
(Eqs. (\ref{eq:basurana})), and disk scale factors are 2.7 kpc and 300 pc.
Fig. \ref{fig:mvszcmpre} shows for values of $\theta=-20^\circ$ and
$-65^{\circ}$ that the two means of determining 
the distribution yield results that differ by amounts that are at most
comparable to the error estimates in the Monte Carlo integrals. This
gives one some confidence in the correctness of the procedure.  The
first form of the distribution has a much simpler integrand and
requires about a fourth as much computer time,  so it will be used
hereafter in all calculations. The 189 days and 12.6 million
source stars used to scale the
ordinate are the span of the MACHO (MAssive Compact Halo Objects) bulge
observations for the 1993 bulge season (Alcock {\it et al} 1996) and
the number of sources they monitored. All of the time scale frequency
distribution curves in the figures are such that $\int_{t_{{\s
E}1}}^{t_{{\s E}2}}F(t_{\s E})dt_{\s E}$ is the number of events with
$t_{{\s E}1}<t_{\s E}<t_{{\s E}2}$ for the 12.6 million source stars over
the 189 day period. 
Alcock {\it et al} (1996) empirically determine a sampling efficiency
of detection as a function of event time scale, and they further
reduce this efficiency by a factor of 0.75 for main sequence turnoff
stars to account for the effects of blending. There is no reduction  
of the sampling efficiency assumed for the clump giant sources.
We use the reciprocal of the overall efficiencies as the inferred number
of events of a particular time scale that would have been detected had
the overall efficiency of detection been 100\%. {\it E.g.,} about 7.41 4.4
day events would have been detected instead of one if 100\% efficiency
were possible. The MACHO detections as thereby enhanced are binned
into 5 day intervals, and a histogram of the distribution of the
number of MACHO detections in the 189 day period of observation is
plotted for comparison to the model distributions of time
scale frequencies below. Only the 40 secure events for single stars
are retained in the MACHO data set.

For the remainder of our examples with the exception of
Fig. \ref{fig:ef133065},  we choose the velocities of source and lens to be 
determined by their position in the galaxy. For $r>3.3$ kpc,
we assume the stars have ``disk'' properties, with a flat rotation
curve at 210 km/sec and velocity dispersions represented by
$\sigma_{\s db}=16$ km/sec and $\sigma_{\s d\ell}=20$ km/sec in the
directions of galactic latitude and longitude respectively.  For
$r<3.3$ kpc, the stars have ``bulge'' properties with circular
velocities decreasing linearly from 210 km/sec at 3.3 kpc to zero at
the center of the galaxy, and with velocity dispersions $\sigma_{\s
bb}=\sigma_{b\ell}=110$ km/sec. The dimensionless distances $z$ and
$\zeta$ are chosen randomly from their respective intervals and 
their values determine the velocity properties of the lens and source
respectively. For example, both lens and source could lie in the bulge
with high velocity dispersions for both. We shall also assume the Zhao
(1996) bulge model described above ($M=2.2\times 10^{10}M_{\s\odot}$)
along with a double exponential disk model with scale lengths of 2.7
kpc in the plane and 300 pc perpendicular to the plane for all cases
below.  We shall truncate the inner part of the otherwise singular
nucleus in the Zhao model by requiring the stellar mass density to depend
only on $z^{\s\prime}$ for values of $r=\sqrt{x^2+y^2}<560$ pc. This
avoids the very large event rates and optical depths as lines of sight
pass near the otherwise singular density at the galactic center. 

 Our choice of the $z$ and
$\zeta$ dependence of the stellar velocities is somewhat arbitrary,
although the rotation of the inner galaxy is close to the rigid body
rotation of the bar pattern adopted by Zhao (1996), and the flat
velocity curve of the disk is that determined by Clemens (1985) and
Alvarez {\it et al} (1990).  However, the latter two sets of authors find the
flat rotation curve to apply for $r>2$ kpc with gas cloud velocities
even increasing at smaller distances. These high central velocities
may not be circular, however. The $z$ and $\zeta$ dependence
of the velocities used here is perhaps the simplest possibility that is
semi-consistent with the observations, although more complicated
(still unmotivated) dependences could be handled in principle. 

Six mass functions that will be considered are shown below.
Proper overall coefficients are introduced in the integrals to yield the
Bahcall and Soneira (1980) local density of $0.05M_\odot/{\rm pc^3}$
for the disk and a bulge mass of $2.2\times 10^{10}M_\odot$. 

\noindent Basu \& Rana
\begin{eqnarray}
\phi(m)&=&1.2588m^{-1}\qquad 0.08\leq m\leq 0.5327,\nonumber\\
       &=&0.3527m^{-3}\qquad 0.5327\leq m\leq 1.205,\nonumber\\
       &=&0.7294m^{-6.83}\qquad 1.205\leq m\leq 2.0, \label{eq:basurana}
\end{eqnarray}
\noindent Modified Basu \& Rana
\begin{eqnarray}
\phi(m)&=&0.1919m^{-2.5}\qquad 0.08\leq m\leq 0.5327,\nonumber\\
       &=&0.1401m^{-3}\qquad 0.5327\leq m\leq 1.205,\nonumber\\
       &=&0.2861m^{-6.83}\qquad 1.205\leq m\leq 2.0, \label{eq:basumdifd}
\end{eqnarray}
\noindent Holtzman {\it et al}
\begin{eqnarray}
\phi(m)&=&0.7795m^{-1}\qquad 0.08\leq m\leq0.7,\nonumber\\
&=&0.5081m^{-2.2}\qquad 0.7\leq m\leq 2.0, \label{eq:holtzman}
\end{eqnarray}
\noindent Modified Holtzman {\it et al}
\begin{equation}
\phi(m)=0.2542m^{-2.2}\qquad 0.08\leq m\leq2.0 \label{eq:modholtzman}
\end{equation}
\noindent Kroupa
\begin{eqnarray}
\phi(m)&=&0.1038m^{-2.35}\qquad 0.075\leq m\leq 0.35,\nonumber\\
       &=&0.6529m^{-.6}\qquad 0.35\leq m\leq 0.6,\nonumber\\
       &=&0.2674m^{-2.35}\qquad 0.6\leq m\leq 10.0, \label{eq:kroupa}
\end{eqnarray}
\noindent Gould, Bahcall, \& Flynn as extended by Mera {\it et al}
\begin{eqnarray}
\phi(m)&=&4.55\times 10^{-4}m^{-4}\qquad 0.08\leq m\leq 0.11,\nonumber\\
       &=&0.8231m^{-0.6}\qquad 0.11\leq m\leq 0.6,\nonumber\\
       &=&0.3371m^{-2.35}\qquad 0.6\leq m\leq 10.0. \label{eq:gould}
\end{eqnarray}
The mass functions are such that $m\phi(m)dm$ represents the mass of
stars/${\rm pc^3}$ in mass range $dm$ about $m$  normalized such
that the total mass density is $1M_\odot/{\rm pc^3}$.
The first mass function (Eq. \ref{eq:basurana}) is that of Basu and
Rana (1992) for local stars, and the second (Eq. (\ref{eq:basumdifd}))
is the same except the index on the low mass region is changed from -1
to -2.5.  The third mass function Eq. (\ref{eq:holtzman})  is determined
by (Holtzman {\it et al}, 1998), and the fourth
(Eq. (\ref{eq:modholtzman}) is our modification, where we have changed
the index  from -1 to -2.2 in the low mass region.
The fifth and sixth mass functions (Eqs. (\ref{eq:kroupa}) and
(\ref{eq:gould})  are those of
Kroupa (1995) and Gould, Bahcall \& Flynn (1997) as represented by
Mera, {\it et al} (1998).  The coefficients are those
appropriate to the lower 
mass cutoff of $0.08M_\odot$ except for the Kroupa mass function which
is cut off at $0.075M_\odot$ for later comparison with some results of
Mera {\it et al}.  We will extend the latter two mass functions into
the brown dwarf region where the coefficients are altered to be
consistent with the normalization of $1M_\odot/{\rm pc^3}$.  The
modification of the Gould {\it et al} mass function by Mera {\it et
al} consists of the top line in Eq. (\ref{eq:gould}) for
$m<0.11$. This modification is not supported by the Gould {\it et al}
data, but we will keep it for the purposes of illustration.  Also, the
index of -0.6 in the second line of Eq (\ref{eq:gould}) is reduced to
about -1.0 if the unresolved binaries are accounted for (A. Gould,
private communication, 1998).  However, we shall maintain the -0.6
index again for illustration. The Kroupa
and the Basu and Rana mass functions are based on measured parallaxes of
the nearby stars, the Gould {\it et al} mass function is
derived from a luminosity function of more distant M stars (Mera {\it
et al} 1998) and the Holtzman mass function is  determined
for bulge stars within Baade's window.   The
Basu and Rana mass function and more often its modified form will be
used in most of the following examples

\begin{figure}[ht]
\plotone{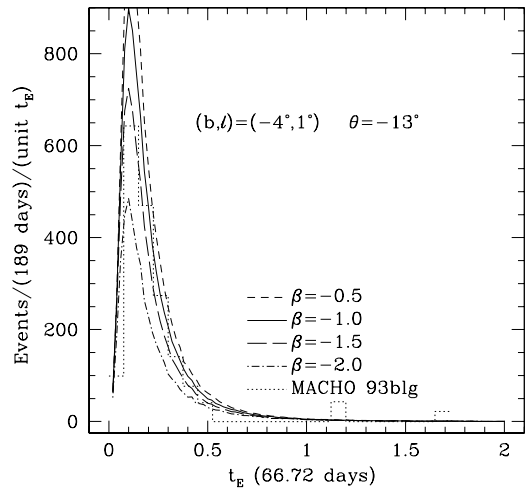}
\caption{Effect of changing the source cumulative luminosity function
index $\beta$. The decrease  
in the number of events as the magnitude of the index increases is as
expected since fewer sources are visible.  Zhao (1996) bar and nucleus
model with bulge mass $=2.2\times 10^{10}M_\odot$. Double exponential
disk model with scale lengths of 2.7 and 0.3 kpc in the disk and
perpendicular to the disk with local value of $\rho_{\s m}=0.05M_{\odot}/{\rm
pc^3}$.  $v_{\s L0},v_{\s S0}=-210$ km/sec for
$z,\zeta<0.5875,\;=2.424(z, \zeta-1)$ for $z,\zeta>0.5875$. Modified
Basu and Rana mass function ($0.08<m<2.0$). $(\sigma_{\s db},\sigma_{\s
d\ell},\sigma_{\s bb},\sigma_{b\ell})=(16,20,110,110\;{\rm
km/sec})$. The ordinate corresponds to 12.6 million source stars monitored
for 189 days. \label{fig:efbeta}}   
\end{figure}

\begin{figure}[ht]
\plotone{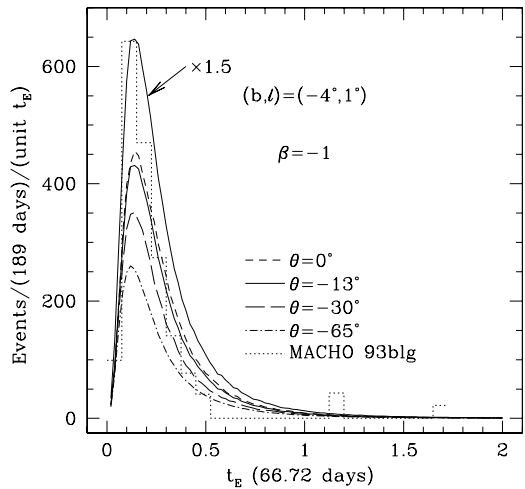}
\caption{Effect of changing the inclination of the bar axis to the
line of sight. The model is the same as that in Fig \ref{fig:efbeta} 
except the unmodified Basu and Rana mass function is
used. \label{fig:ef0133065}} 
\end{figure}

\begin{figure}[ht]
\plotone{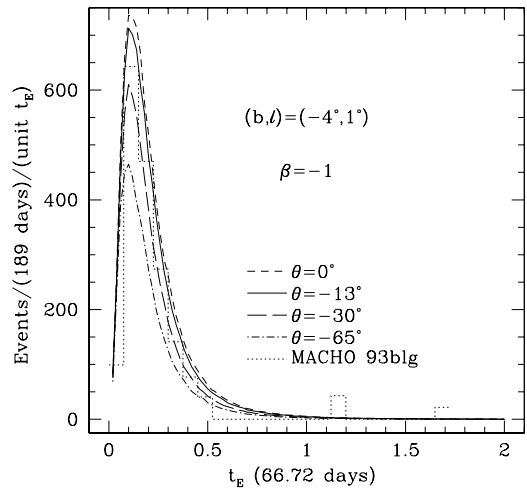}
\caption{Effect of changing the velocity model. Galactic model is the
same as Fig \ref{fig:ef0133065} except $v_{\s L0}=210$, $v_{\s
S0}=0$ km/sec, $\sigma_{\s Lb}=\sigma_{\s L\ell}=\sigma_{\s
Sb}=\sigma_{\s S\ell}=110$ km/sec. Short time scale event rate and
total number of events is increased significantly. \label{fig:ef133065}}  
\end{figure}

\begin{figure}[ht]
\plotone{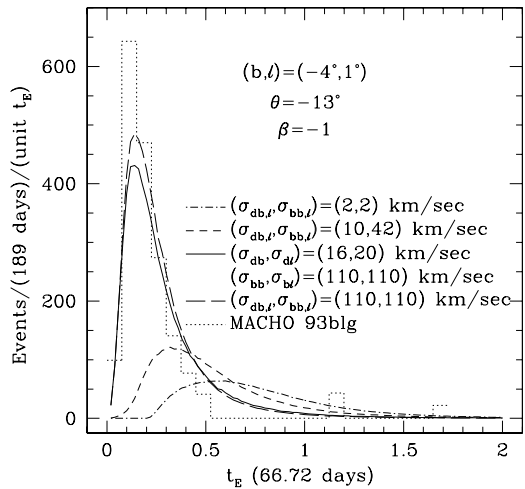}
\caption{Effect of changing the velocity dispersions.  The
model is the same as that in Fig. \ref{fig:efbeta} except for the
changed velocity dispersions, the fixed value of $\beta$ and the use
of the unmodified Basu and Rana mass function. \label{fig:veldispers}}   
\end{figure}

\begin{figure}[ht]
\plotone{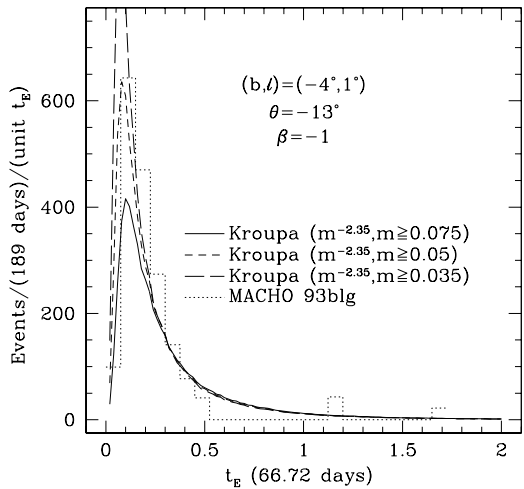}
\caption{Comparison of the time scale frequency distributions obtained
from the Kroupa mass functions extended into the Brown dwarf region
with that from the MACHO 1993 bulge data.  All model parameters except
the mass function  are those of
Fig. \ref{fig:efbeta}. \label{fig:efmassfns}} 
\end{figure}

\begin{figure}[ht]
\plotone{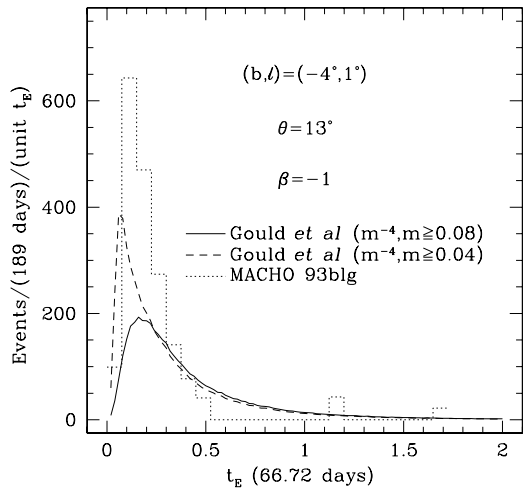}
\caption{Same as Fig. \ref{fig:efmassfns} except Gould {\it et al}
mass functions are used instead of the Kroupa mass functions.
Note, however, that the extension of the Gould
{\it et al} mass function into the brown dwarf region by Mera {et al}
(1998) is completely speculative.   \label{fig:efmassfns2}}
\end{figure}

\begin{figure}[ht]
\plotone{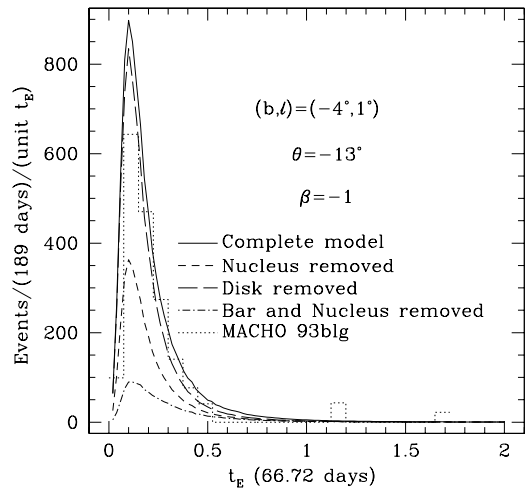}
\caption{Relative contributions of several model constituents to the
model time scale frequency distributions, where the model is that
of Fig. \ref{fig:efbeta}. \label{fig:efremovals}}
\end{figure}

Fig. \ref{fig:efbeta} shows the effect of changing the luminosity
index $\beta$ from -0.5 to -2.0 superposed on the
distribution obtained from the 1993 MACHO bulge data (Alcock {\it et
al} 1996).  (Although we shall see that the modeled distributions
should be averaged over the MACHO field positions for comparison to
the empirical distribution of time scale frequencies, the choice of a
line of sight toward galactic coordinates
$(b,\ell)=(-4^\circ,1^\circ)$ (in Baade's window) yields distributions
very close to this average shown in Fig. \ref{fig:machoave}.)   Fewer
sources are visible as this index gets more negative, and the expected
decrease in the number of events observed as the index varies from
-0.5 to -2.0 is displayed. The modified Basu and Rana mass function is
used, and the curve for $\beta=-1$ is the same as that shown as the
``complete model'' in Fig. \ref{fig:efremovals}. The other model
parameters are specified in the figure caption.  
Fig. \ref{fig:ef0133065} shows the distribution of time scale
frequencies  for several bar orientations using the Basu and Rana mass
function.  The number of events decreases as the magnitude of the
inclination of the bulge increases as expected, since fewer stars are
located along the line of sight as the long axis of the bulge is
pointed further away from the observer. Since the bulge distribution is so
dominant over that of the disk, the time scale frequency distribution
is almost proportional to the mass of the bulge.  Hence, the curve for
$\theta=-13^\circ$ is in some ways a closer match to the MACHO data if
the bulge mass is increased by a factor of 1.5.  Zhao {\it et al} (1996) also
pointed out the sensitivity of the event rate to the bulge mass, but
notice here that simply increasing the bulge mass to match the peak in
the distribution produces too many events in the 8 to 40 day range. None
of the model distributions 
seem to produce as many long time scale events as the MACHO group
observed. 
To show how a different velocity model can affect the model time scale
frequencies, we repeat the exercise
of Fig. \ref{fig:ef0133065}, where all parameters are identical to
those used to construct that figure except the velocities.  Instead of
assigning circular velocities and velocity dispersions as a function
of position in the galaxy, we here assume $v_{\s L0}=210$ km/sec for
all $z$, $v_{\s S0}=0$ and $\sigma_{\s Lb}=\sigma_{\s
L\ell}=\sigma_{\s Sb}=\sigma_{\s S\ell}=110$ km/sec for all $z$ and
$\zeta$.  Although this dispersion 
is not realistic, it might be a reasonable approximation if most of
the lenses were in fact in the bulge.  On the other hand, keeping the
source circular 
velocities zero while assuming that all of the lenses maintain the
flat velocity curve of the disk already introduces a significant
relative transverse velocity between source and lens before the
dispersions are applied.  Then the large dispersions superposed onto this
velocity difference result in a large increase in the relative number
of short time scale events and in the total number of events. The
results shown in Fig. \ref{fig:ef133065} show this large increase in
the time scale frequency distribution for all of the bulge
orientations. Even with the use of the Basu and Rana mass function,
the MACHO data is now well matched by the distribution for
$\theta=-13^\circ$ without increasing the bulge mass, but this match
is an artifact of an unrealistic velocity model.
We return now to the velocity model of Fig. \ref{fig:ef0133065} where
both circular velocities and the dispersions depend on $z$ and $\zeta$
along with the other parameters assumed in the construction of that
figure to show explicitly the dependence of the model time scale 
frequency distribution on velocity dispersion. Fig. \ref{fig:veldispers}
displays the distributions for several choices of velocity
dispersions, where more long time scale events are
produced but with fewer total events, and the peak in the distribution
moves to longer time scales as the velocity dispersion is decreased.
Notice, however, that increasing the velocity dispersion of the disk
part of the galaxy to the values appropriate to the bulge results in
only a modest increase in the distribution over that where only the
bulge has the high dispersion.  This is because in the latter case
most of the lenses already have a large dispersion, since relatively
few lenses are in the disk distribution.
The sensitivity of the microlensing time scale frequency distribution
to the stellar mass function of the lenses is demonstrated in
Figs. \ref{fig:efmassfns} and \ref{fig:efmassfns2}, where consequences
of assuming the several mass functions given in
Eqs. (\ref{eq:basurana}) to (\ref{eq:gould}) for various extensions
into the brown dwarf region are compared with the MACHO data. 
In Fig. \ref{fig:efmassfns} we show the time scale frequency
distributions with $\theta=-13^\circ$ for 
three Kroupa mass functions with successively larger
excursions into the brown dwarf region as indicated by the minimum
mass. The indices shown in the figure apply only for the lowest mass
range in the definitions of the mass functions.  Except for the mass
functions, all model parameters are those of Fig. \ref{fig:efbeta}
with $\beta=-1$. Fig. \ref{fig:efmassfns2} shows the
distributions of the Gould {\it et al} mass functions, where the same
model assumptions apply as in Fig. \ref{fig:efmassfns}.  
First notice the degree of sensitivity of the model time scale
frequency distributions to the mass functions, the only variation in
the models. Herein lies the eventual power of the microlensing
technique to constrain the galactic mass function far from the Sun.
The Kroupa mass function extended down to $m=0.05$ yields a
distribution that is a fair match to the MACHO data, although it is
somewhat ``thin'' in the region around the peak.  Improving the fit
on the longer time scale side of the peak by increasing the galactic
bulge mass makes the fit worse for $t_{\s E}$ near 33 days. A similar
problem is faced by the Basu and Rana mass function for the bulge mass
increased by a factor of 1.5 in Fig. \ref{fig:ef0133065}, where there
are too many events in the 8 to 40 day range of $t_{\s E}$.  Perhaps
the best fit to the MACHO data is for the modified Basu and Rana mass
function in Fig. \ref{fig:efbeta} for $\beta=-1$ and
$\theta=-13^\circ$ even though the peak is somewhat too high. This fit
could even be improved somewhat by 
decreasing the mass of the bulge slightly (this would lower the peak),
while making the -2.5 index in the M star region slightly less
negative (this would ``fatten'' the curve to bring it back into
coincidence with the Macho data. It is noteworthy that the best fit to
the MACHO data is obtained with only hydrogen burning stars, albeit
with the M star region enhanced with an index of -2.5.

Mera {\it et al} (1998), with a different galactic model, find a time
scale frequency distribution that 
is consistent with the MACHO data using the Kroupa mass function
with lower bound $0.075M_{\odot}$---slightly into the brown dwarf
region. This same mass function yields a poor match
to the MACHO data with the model used
here. This illustrates the model  and perhaps procedure dependence of
conclusions about the mass functions when so may crucial galactic
parameters are so poorly constrained by observations.  

\begin{figure}[ht]
\plotone{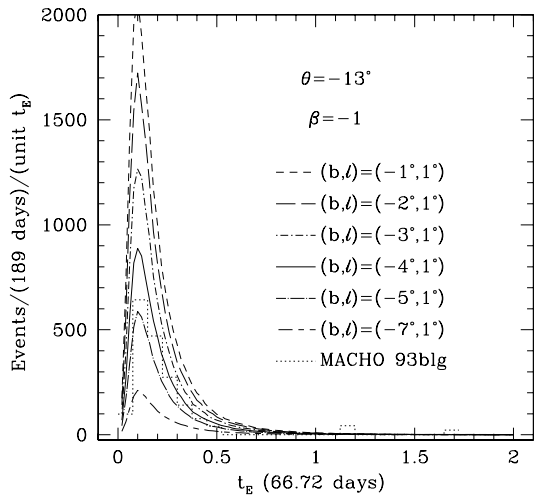}
\caption{Dependence of the time scale frequency distribution on
the galactic latitude of the line of sight with the longitude fixed at
$1^\circ$. The singularity in the
nucleus density distribution is  removed as discussed in the
text. Model is that of Fig. \ref{fig:efbeta}.\label{fig:eflatitude}}
\end{figure}

\begin{figure}[ht]
\plotone{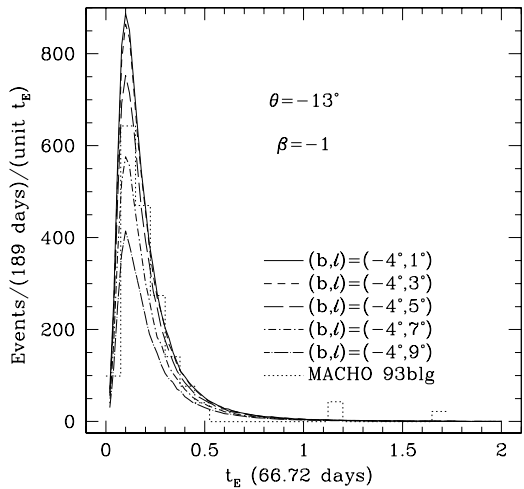}
\caption{Same as Fig. \ref{fig:eflatitude} but longitude is varied at
constant $-4^\circ$ latitude\hspace*{1.05in}. \label{fig:eflongitude}}
\end{figure}

\begin{figure}[ht]
\plotone{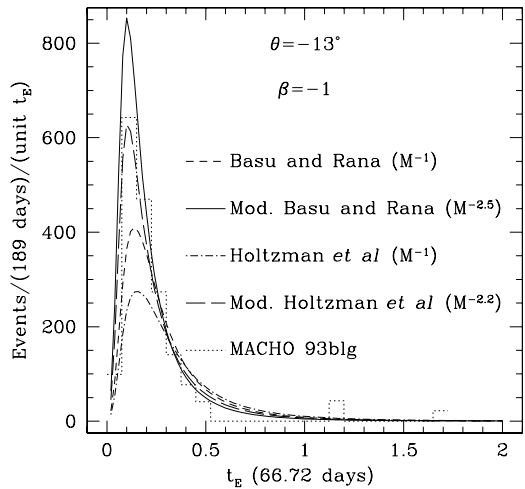}
\caption{Model time scale frequency distributions averaged over the
lines of sight to the 24 MACHO fields for modified and unmodified
Basu and Rana mass functions and modified and unmodified Holtzman {\it
et al} mass functions.  The variation of the mass functions in the M
star region is indicated in parentheses. The  model is that of 
Fig. \ref{fig:efbeta} except for the coordinates of the lines of
sight and the explicit mass functions. \label{fig:machoave}}
\end{figure}

Extending the Kroupa mass functions further into the brown dwarf region to
enhance the time scale frequency distribution is only partially
successful.  If one believes the 
statistics of small numbers characterizing the MACHO data, increasing
the fraction of the mass distribution in the brown dwarfs makes the
distribution too narrow and shifts the peak too far toward shorter
time scale events.  The distributions so derived have (slightly) too
few events of moderate time scales.  The Gould {\it et al} mass functions,
(Fig. \ref{fig:efmassfns2}) are
even worse matches to the MACHO data as also found by Mera {\it et al}
for their model.  This mass function has many fewer stars between 0.11
and $0.35M_\odot$, and for $m>0.08$ there is only a short range where the
mass function appears to turn up steeply in the Mera {\it et al}
extension. The severe lack of  
small mass stars leads to the shift in the peak in the distribution to
longer time scales, more moderately long period events, and a rather
modest total number of events. Extending the very steep mass function
into the brown dwarf region ($m>0.04$) fails worse than the same
exercise with the Kroupa mass function in matching the MACHO data
since the peak of the distribution is moved even further toward very
short time scales ($<$ 4.5 days), exceeding the MACHO numbers there while
failing by a factor of at least two to produce a sufficient number of
events of time scales between 5 and 20 days.  Since the modified Basu
and Rana mass function gives the best fit to the MACHO
data, we shall use this mass function in subsequent examples, except
in Fig. \ref{fig:machoave} where the modified Holtzman {\it et al}
mass function is shown to also give a reasonably good fit to the MACHO
data. 
Fig. \ref{fig:efremovals} shows the relative importance of various
contributions to the model mass distribution in determining a time
scale frequency distribution for $\theta=-13^\circ$. Removing the disk
stars has relatively little effect on the distribution confirming the
suspicion that most of the lenses are in fact located in the
bulge. Consistently, the number of events is drastically reduced when
only the disk is 
present, although the peak in the distribution is not changed much
because the velocities still have the $z$ and $\zeta$ dependence of
Figs. \ref{fig:efbeta} and \ref{fig:ef0133065}. Removing the nucleus decreases the
number of events by a factor of about 2.5.  The nucleus contributes
about 21\% of the total bulge mass, but has such a large effect
because the line of sight to $(b,\ell)=(-4^\circ,1^\circ)$ for
$\theta=-13^\circ$ encounters a comparable number of nucleus and bar
stars.  We should point out here that Zhao {\it et al} (1996) are able
to get a good match to the MACHO data with a comparable mass function
and velocity dispersion as that used here without the nucleus, whereas we
require the nucleus be kept for a good match.  Like the Mera {\it et
al} (1998) match to the MACHO data with the Kroupa mass function which
failed for our model, this again illustrates the model and procedure
dependence of estimated time scale frequencies that is independent of
and thereby weakens the constraints on the mass function.
The Zhao model is also very sensitive to the galactic coordinates of the
line of sight.  This is shown in Figs. \ref{fig:eflatitude} and
\ref{fig:eflongitude}, which show the time scale frequency distributions
along two perpendicular traces through Baade's window whose galactic latitude
($-4^\circ$) bisects the more extended region where the MACHO dat was
obtained. The smaller scale lengths assumed in the z direction for
both the  bar and nucleus results in the more rapid fall off in events
at higher galactic latitudes than for higher galactic longitudes. Recall
that we have eliminated the singular nature of the nucleus by
truncating the nucleus density at the galactic center by assuming that density
uniform for constant $z^{\s\prime}$ inside 560 pc.  The nucleus thereby loses
about 36\% of its mass but the whole bulge loses only 7\%.  If the
nucleus is not truncated, both the rate of events and the optical
depth become very large as the line of sight approaches the center of
the galaxy.  Zhao {\it et al} (1996) avoid this problem by eliminating
the nucleus altogether, but we require the nucleus stars to match the
MACHO data as discussed above.  

The large dependence of the event rate
on coordinates of the line of sight means that microlensing will also
eventually prove a powerful constraint on bar model star distributions
as well as mass functions if other parameters such as the velocity
dispersions can be adequately defined.  As the MACHO fields are
spread over several degrees near Baade's window, (Fig. 1 of Alcock
{\it et al} (1996)), it is appropriate that the model event
frequency distributions be averaged over the MACHO fields for
comparison with their results.  This average is shown in
Fig. \ref{fig:machoave} for Basu and Rana and Holtzman {\it et al}
mass functions  and 
their modifications by the more negative indices for low mass stars. Both
modified mass functions yield good agreement with 
the data with the best match probably obtained with an index between
-2.2 and -2.5.  However, given all of the uncertainties in the model
parameters discussed above and the simplicity of the model itself,
this agreement must be regarded as fortuitous and should not be
regarded as a definitive constraint on the mass function.  Note,
however, that there is as yet no need to assume that there is any
significant fraction of the bulge mass in brown dwarfs---a conclusion
also reached by Zhao {\it et al} (1996). On the other hand, we do need
more M stars in the mass functions than obtained from the star counts
in order to match the MACHO data.  There may be a not unreasonable
galaxy model that will yield a sufficient number of short time scale
events with a $m^{-1}$ variation of the mass function in the M star
region, but the necessary changes in the parameters of the model
adopted here seem too extreme.  But the uncertainties in the most
relevant mass function of the galactic bulge stars obtained by
Holtzman {\it et al}, the paucity of the microlensing data and
modest observational constraints on the galactic model itself do not
yet warrant a search for a drastically different galactic model.
We now turn to a determination of the optical depth as a function of
the line of sight.  We shall see that a similar average over the MACHO
lines of sights for our best matching model also gives reasonable 
agreement with the MACHO estimates. We shall eventually conclude that
the Zhao model distribution of stars with our assumptions about the
mass function and the velocity distribution yields a good match
to both the estimated optical depth and the time scale frequency
distribution of the 1993 MACHO data, but that the meagerness of this
data and the uncertainties within the galactic models means that the
microlensing constraints on both the mass function and the
distributions of lens and source stars are very weak indeed. 

\section{Optical depth}

If we use the definition of a  microlensing event as a source at
distance $D_{\s OS}$ passing within the Einstein ring radius $R_{\s
E}$ of a lens at distance $D_{\s OL}$, the cross section of the lens
for such an event is just $\pi R_{\s E}^2$, where $R_{\s E}(m,D_{\s
OL},D_{\s OS})$ is given by Eq. (\ref{eq:re}). For such lenses within a
slab of thickness $dD_{\s OL}$ at $D_{\s OL}$, the probability $dP$ that a
ray from a source will pass through an Einstein ring on its way to the
observer is thus $\pi R_{\s E}^2(dn_{\s L}(m,D_{\s OL})/dm)dm\,dD_{\s
OL}$, where $n_{\s L}$ 
is the number density of lenses and where we have selected only the
lenses with masses within $dm$ of $m$.  If $I$ is the intensity of
rays from the source that have {\it not} passed through an Einstein
ring in traveling distance $D_{\s OS}-D_{\s OL}$ on their way to the
observer, then the fractional change in traversing $dD_{\s OL}$ from
lenses with masses within $dm$ of $m$ is $dI/I=-dP$ or 
\begin{equation}
\frac{dI}{I}=-\frac{4\pi Gm}{c^2}\frac{D_{\s OL}(D_{\s OS}-D_{\s
OL})}{D_{\s OS}}\frac{1}{m}\frac{d\rho_{\s L}(m,D_{\s OL})}{dm}dm\,dD_{\s OL},
\label{eq:dtau}
\end{equation}
where $\rho_{\s L}$ is the stellar mass density and Eq. (\ref{eq:re}) has
been used.  Integration of Eq. (\ref{eq:dtau}) leads to the optical
depth for microlensing to distance $D_{\s OS}$ of 
\begin{equation}
\tau=\frac{4\pi G}{c^2}\int_0^{D_{\s OS}}\rho_{\s L}(D_{\s OL})\frac{D_{\s
OL}(D_{\s OS}-D_{\s OL})}{D_{\s OS}}dD_{\s OL}, \label{eq:tau}
\end{equation}
where cancellation of the masses in Eq. (\ref{eq:dtau}) allows
immediate integration over the lens mass function. Finally we
introduce the dimensionless variables used earlier and average the
optical depth over the distribution of visible sources along the line
of sight to obtain the result of Kiraga and Paczynski (1994)
\begin{eqnarray}
\left<\tau\right>&=&\frac{4\pi G D_{\s 8}^2}{c^2}\int_{0.1}^{1.2}\!\!
\int_0^\zeta \rho_{\s L}(z)\frac{z(\zeta-z)}{\zeta}\times\cr \nonumber
&&n_{\s s}(\zeta)
\zeta^{2+2*\beta}dz\,d\zeta{\mbox{\Huge /}}\!\!\!\int_0^{1.2} n_{\s s}(\zeta)
\zeta^{2+2\beta}d\zeta, \label{eq:tauave}
\end{eqnarray}
where $n_{\s s}(\zeta)$ is the source number density and where the
coefficient is $3.86\times 10^{-5}$ if $\rho_{\s L}$ is expressed in
$M_{\s\odot}/{\rm pc^3}$. We shall omit the brackets around $\tau$
hereinafter, but always understand that it is averaged over the source
distribution.

\begin{figure}[ht]
\plotone{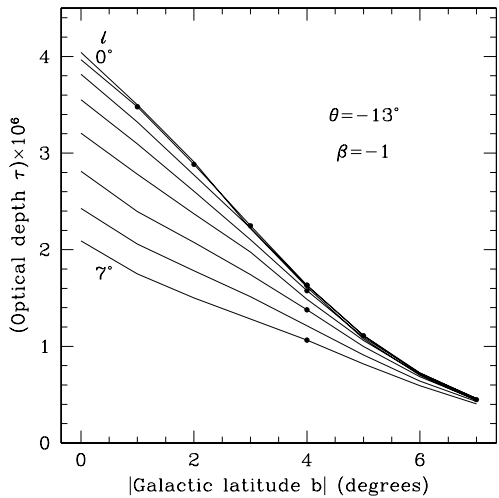}
\caption{Dependence of optical depth on direction of the line of sight
for the Zhao bar and truncated nucleus $(M=2.2\times 10^{10}M_\odot)$
with double exponential disk with scale lengths of 0.3 and 2.7 kpc and
local density of $0.05M_\odot/{\rm pc^3}$.  The dots correspond to the 
directions for which the time scale frequency distribution was
determined in Figs. \ref{fig:eflatitude} and  \ref{fig:eflongitude}.
\label{fig:tautotz}}
\end{figure}

\begin{figure}[ht]
\plotone{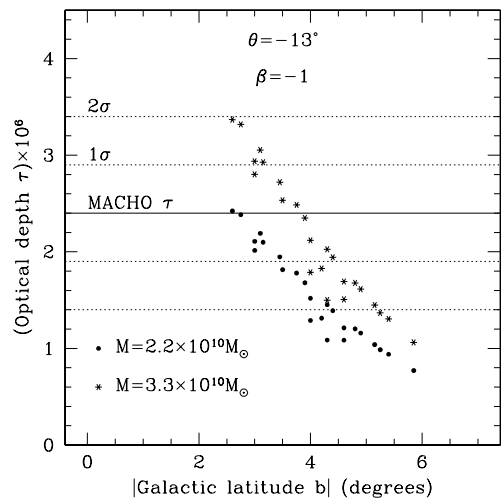}
\caption{Comparison of model optical depths for the MACHO fields with
the empirically determined average value. The model is the same as
that in Fig. \ref{fig:tautotz} except for the specific bulge masses
indicated for the two sets of points. \label{fig:taumacho}}
\end{figure}

\begin{figure}[ht]
\plotone{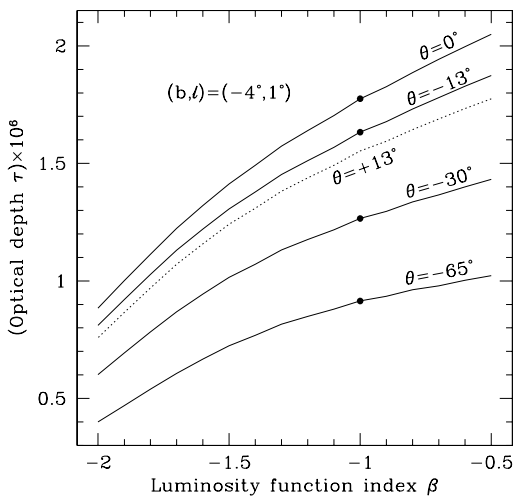}
\caption{Variation of optical depth with inclination of the bar axis
and with the source luminosity function index.  The model is the same as in
Fig. \ref{fig:tautotz} except for the indicated variations in $\theta$
and $\beta$. The dots correspond to the
values of $\theta$ and $\beta$ for which time scale frequencies were
determined in Fig. \ref{fig:ef0133065}. \label{fig:taubeta}}
\end{figure}

\begin{figure}[h]
\plotone{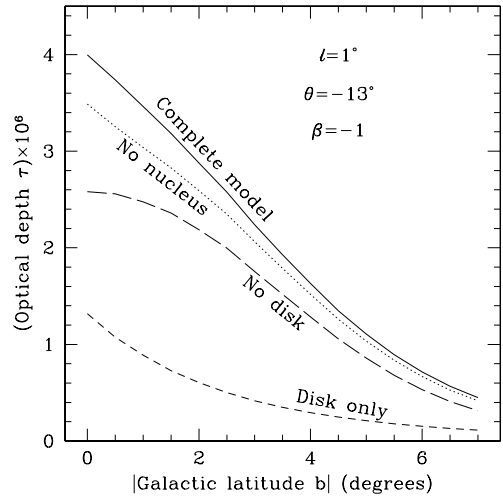}
\caption{Contributions of various parts of the model of
Fig. \ref{fig:tautotz} to the optical depth as a function of galactic
latitude. \label{fig:tauremovals}}
\end{figure}

The integral in the numerator of Eq. (\ref{eq:tauave}) is evaluated
with a Monte Carlo technique like that used for the time scale
frequency distribution, whereas the denominator is evaluated with a
standard numerical technique. Our examples of optical depth will be mostly
limited to the model of Fig. \ref{fig:efbeta}, as that model which
best matches the time scale frequency distribution averaged over the
MACHO lines of sight.  This model is the Zhao bar and truncated
nucleus with total mass of $2.2\times 10^{10}M_\odot$ and the double
exponential disk with scale lengths of 0.3 and 2.7 kpc, but the
optical depth is independent of the mass function.  The optical depth
is quite sensitive to the coordinates of the line of sight for this
model as indicated in Fig. \ref{fig:tautotz}, where we show $\tau$ as
a function of latitude and longitude around Baade's window. The
dots in Fig. \ref{fig:tautotz} mark the coordinates where we have
evaluated the time scale frequency distribution in
Figs. \ref{fig:eflatitude} and \ref{fig:eflongitude}, and they display
the expected correlation between the optical depth and number of
events. If we reduce the radius inside of which the mass density of
the nucleus is constant, the optical depth, like the time scale
frequency distribution, is increased  for latitudes and longitudes
less than about $4^\circ$, but is little affected for larger
coordinates. 
In Fig. \ref{fig:taumacho} optical depths for the MACHO line of sight
directions 
are distributed across horizontal lines designating the estimated
MACHO empirical optical depth averaged over all of the 24 fields and 
1 and $2\sigma$ error estimates.  The points for a bulge mass of
$2.2\times 10^{10}M_{\s\odot}$, that which gave such a good match for
the time scale frequency distribution, fall somewhat low with about
half falling within the $2\sigma$ lines. In some sense, this can be
regarded as not unreasonable agreement given the uncertainties in the
galactic model and the sparseness of the data.  Recall in
Fig. \ref{fig:ef0133065} that we multiplied the time scale
frequency distribution obtained for the Zhao model bulge together with
our standard velocity model but with the Basu and Rana mass function  
by a factor of 1.5 to match the peak in the empirical
distribution. Although this leads to too many mid duration events, it
could be regarded as a better fit.  This multiplication simulates a
similar increase in  the bulge mass. In Fig. \ref{fig:taumacho}, we
see that the optical depths for the MACHO field coordinates are more
compatible with the empirical estimate for the higher mass
bulge. However, in
addition to the poorer match of the Basu and Rana mass function to the
moderate duration events obtained by MACHO in
Fig. \ref{fig:ef0133065},  the higher mass bulge may
be less compatible with the stellar kinematics as determined by Zhao
(1996). The model optical depth averaged over the MACHO field
directions is $1.54\times 10^{-6}$ for a bulge mass of $2.2\times
10^{10}M_{\s\odot}$ and $2.14\times 10^{-6}$ for $3.3\times
10^{10}M_{\s\odot}$ compared with $(2.4 \pm 0.5)\times 10^{-6}$ for
the MACHO empirical value. 
In Fig. \ref{fig:taubeta} we show the effect of different positions of
the bar axis along with luminosity function index $\beta$ for the direction
$(b,\ell)=(-4^\circ,1^\circ)$. The optical depth shows the expected
decrease as the magnitude of the index increases---location of fewer
visible sources at the larger distances pushes the mean position of
the sources closer to the observer with less optical depth
resulting. The optical depth also shows the expected decrease as the
bar is rotated further from the line of sight.  We show the optical
depth for the near side of the bar in the fourth quadrant
($\theta=+13^\circ$) to demonstrate the expected decrease when the
near side of the bar is not in the same quadrant as the line of
sight. The dots indicate values of $\beta$ and $\theta$ for which we
determined the time scale frequency distribution in
Fig. \ref{fig:ef0133065} where the optical depths and the number of
events are again appropriately correlated.
As a companion to Fig. \ref{fig:efremovals} showing the relative
contributions of various parts of the galactic model to the time scale
frequency distribution, we show in Fig. \ref{fig:tauremovals} the
effect of the same dissection of the model on optical depth. Again,
the disk contributes relatively little to the optical depth---like its
meager contribution to the time scale frequency distributions.  The
nucleus becomes more important when the line of sight passes very
close to the galactic center. The optical depth for the complete model
would turn up sharply at small galactic latitudes in
Fig. \ref{fig:tauremovals} without the truncation of the nucleus.
\section{Discussion}
We have developed a procedure for determining model time scale
frequency distributions for microlensing events toward the center of
the galaxy that can be easily modified for improvements in galaxy
models as they become available. Two methods of evaluating the
integral used to accumulate all of the combinations of variables
leading to the same time scale give the same distribution within the
errors of the Monte Carlo integrations (Fig \ref{fig:mvszcmpre}). This
robustness of the model distribution gives one some confidence that the
computational procedure is sound.  That changes in the distributions
as various parameters are varied in a series of examples were as
expected increases this confidence.  

The complete variable dependence of
the event time scale is maintained within the integral used to
accumulate all the combinations of variables leading to the same time
scale.  This contrasts to some work in the literature where event
rates are determined for all of the lenses being $1M_{\s\odot}$ with a
later convolution over the mass function (e.g. Kiraga and Paczy\'nski
1994; Zhao {\it et al} 1996). However, it seems implicitly assumed in
this latter procedure that galactic models are secure enough that the
mass function can be separated out as the major unknown and
meaningfully constrained by substituting various forms into the
convolution. There would also appear to be some danger of introducing
some artifacts into the computation by such piecewise averaging.
Within the Zhao (1996) bulge model (bar plus nucleus) and the double
exponential Bahcall and Soneira disk model that we have adopted,
the time scale frequency distributions are very sensitive to the total 
bulge mass, the circular velocity model, velocity dispersions and the
direction of the line of sight as
well as the mass function.  Other bulge and disk distributions of
stars will introduce still other 
variations. So until the properties of the galaxy are better
constrained, it seems premature to assign the highest priority to the
mass function in matching the model time scale frequency distributions to
the observations.

With our nominal adornments to the Zhao bulge model, we obtain a
good match to the MACHO time scale frequency 
distribution with an optical depth that is only slightly too low
(Figs. \ref{fig:machoave} and \ref{fig:taumacho} ). But we do this only
by changing the index of either the Basu and Rana or Holtzman mass
function from -1 to -2.5 and -2.2 respectively in the M star region
(Fig. \ref{fig:machoave}). The Kroupa mass function  
has a slope of -2.35 for the latest M stars, but the Gould {\it et al}
mass function is even flatter than that of Basu and Rana (-0.6 vs
-1.0), with the steep extension into the brown dwarf region by Mera {\it et
al} (1998) being completely speculative.  From Figure
\ref{fig:machoave} one is led to an index somewhere between -2.2
and -2.5 for an even better fit to the MACHO data within the overall
model.  But the Holtzman {\it et al} mass function for bulge stars in
the M star region ($\sim m^{-1}$) where most of the lenses reside is
like that of Basu and Rana and also Gould {\it et al} if unresolved
binaries are accounted for in the latter. So the only observational
support for the steeper mass function in the M star region is
consistency with a meager MACHO data set.
Still, microlensing may be detecting M stars that are not
accounted for in the star counts because of difficulties in estimating
the completeness of the optical surveys (Holtzman, private
communication, 1998), or the meagerness of the microlensing data set
may make the empirical event rates not entirely representative.  

With a different
velocity model, or with other than a Zhao bulge distribution of stars,
our match may become less striking.  Still, it is significant that we
found a model consistent with the microlensing data without resorting
to a large population of brown dwarfs as advocated by several
authors ({\it e.g.} Han and Gould (1996); Han 1998).  In fact, we
find that extending the Kroupa and Gould {\it et al} mass functions
into the brown dwarf region produced too many very short time scale
events and an insufficient number of moderate time scale events
(Figs. \ref{fig:efmassfns} and \ref{fig:efmassfns2}), although at
least the Kroupa mass function including brown dwarfs might have been
a better fit if we had just decreased the velocity dispersions or
changed the bulge mass. 
 
Our use of the Kiraga and Paczy\'nski luminosity function index
$\beta$ as a measure of source visibility is useful for verifying
consistent behavior of the distributions derived by this procedure
(Fig. \ref{fig:efbeta}),  but it is
actually not a good representation.  First, the luminosity function is
not so simply behaved in the visible region, and second, blending is a
much more important limitation on source visibility than apparent
brightness.  The representation might be useful in accounting for
source visibility in highly obscured regions, but there are likely to be
too few visible sources in these circumstances to motivate an
extensive microlensing search. Penetration of these regions by using
infrared wavelengths would return us to blending as the most important
limitation.  Although there is a need for a more realistic  model of source
visibility, it is probably less important than other uncertainties
in the galactic model.

The change in the distribution function as we altered the inclination
$\theta$ of the bar axis to the observer-galactic center line also showed
consistent behavior for the model (Fig \ref{fig:ef0133065}) as did the
optical depth (Fig. \ref{fig:taubeta}).  Although
this inclination is uncertain to some extent, which fact motivated our
investigation of its effect, changing the inclination also requires changing
the shape of the bar to remain consistent with the COBE observations
(Zhao 1998).  We left the bar shape unaltered during this
exercise, but other uncertainties preclude a more complicated model.

It not clear just how well our velocity model represents the real
world, but a more complicated one is not warranted now.  Certainly,
the velocity dispersion must have a more complicated $z$ and $\zeta$
dependence than the one that we used mostly ({\it e.g.}
Figs. \ref{fig:efbeta} and \ref{fig:ef0133065}), and the flat circular
velocity of the disk and ramped circular velocity of the bulge can
only be rather crude approximations.  Still, it is useful to realize
the extreme variation in the calculated time scale frequency
distributions simply by changing the velocity dispersions within the
model (Fig. \ref{fig:veldispers}) or by a change in the model itself
(Fig. \ref{fig:ef133065}).  A velocity model, both circular velocities
and dispersions or even more complicated models accounting for the box
like orbits in the Zhao (1996) bulge model, that does not represent
the real galaxy very well may lead to, say, the adoption of a mass
function in matching microlensing data that is far from reality.  

We have so far in this discussion concentrated on the uncertainties in
deducing galactic properties from microlensing time scale frequency
distributions and optical depths toward the galactic center. However,
the sensitivity of such microlensing data to changes in the model
illustrated by our series of examples shows just how powerful the
technique will eventually prove to be in the study of galactic
structure. Figs. \ref{fig:eflatitude}, \ref{fig:eflongitude}, 
\ref{fig:tautotz} and \ref{fig:taumacho} show the very rapid variation
in both optical depth 
and time scale frequency distributions with relatively small changes
in the line of sight direction.  If such variations do not emerge with
the collection and analysis of many more events to obtain the
necessary spatial resolution, one would have to drastically revise
current ideas about the nature of the bar.  Much better spatial
resolution may already be possible from the more than 200 additional
events that have been monitored since 1993, but that remain
unanalyzed and unpublished.  Recall also that all ways to improve the
fit to the data are not equivalent.  In Fig. \ref{fig:ef0133065}, the
Basu and Rana mass function led to a time scale frequency distribution
that produced far too few short time scale events for
$\theta=-13^\circ$, one of the preferred inclination angles for the
bar axis. We increased the number of these events for this inclination
by increasing the galactic mass by a factor of 1.5, an increase that
might be tolerated by other observational constraints. If the MACHO
distribution is close to the correct one, the number of events
predicted between about 15 and 40 days is now too large. A better way
to match the data was to increase the magnitude of the index in the M
star range of the Basu and Rana or Holtzman {\it et al} mass functions
as shown in Fig. \ref{fig:machoave}. 

So microlensing will place constraints on the
galactic distribution of stars, their distributions of velocities and
the mass function, but other more conventional observations must also
be accumulated to independently constrain the parameters for a gradual
convergence to the true properties of the galaxy.  Certainly with only
40 data points, very uncertain velocities and still uncertain star
distribution in spite of the success of the Zhao (1996) model (We have
not even discussed deviations from axial, radial or bar symmetry.), it
is premature to assert any constraints on galactic properties in
general from the microlensing data (except for the existence of the
bar) and in particular on the mass
function. We found a mass function that works along with our other
assumptions, but we are unable to assert that it is the right one.

\section{Conclusions}
\parskip=10pt
We end by enumerating our conclusions.

\noindent 1. The procedure developed here appears to be
computationally sound (Fig. \ref{fig:mvszcmpre} and later
consistencies) and provides a convenient 
framework for rapidly determining the predictions for microlensing
surveys of the galaxy.  Position dependence of various galactic
properties such as stellar velocities is relatively easily
incorporated when observations justify such detail.   

\noindent 2. The sensitivity of the predictions of our overall model
to changes in various galactic parameters used in the model, means
that microlensing will be a major player in the set of many types of
observations that will ultimately constrain the structure, content and
dynamics of the galaxy.

\noindent 3. There is no motivation for invoking a large population of
brown dwarfs to account for the number of short time scale events
observed by the MACHO group.  Brown dwarfs might be accommodated by
reducing the velocity dispersion, for example, but there appears to be
a sharp fall off in the number of  brown dwarfs as companions to local
solar type stars (Mazeh, Goldberg \& Latham 1998), and there is no apparent
theoretical or observational reason for expecting many brown dwarfs in
the bulge.   

\noindent 4. The sensitivity of the microlensing event rates and
optical depth to the direction of the line of sight for bar models
means it will be a sensitive probe of the star distribution in such
models. 

\noindent 5. The disk stars contribute to a minor fraction of the
microlensing events (Figs. \ref{fig:efremovals} and
\ref{fig:tauremovals}).

\noindent 6. We obtain a better fit to the MACHO data by including the
nucleus in the Zhao (1996) bulge model, albeit we truncated the
singular peak in the density to avoid excessively large optical depths
and event rates for lines of sight near the galactic center.  This
truncation is probably academic in any case as obscuration increases
drastically for lines of sight very close to the galactic center and
no data have been taken there.

\noindent 7.  Our model time scale frequency distributions dance
around the empirical distribution from the MACHO observations as
various parameters are varied, but we are able to get a very good
match to this data from the distributions from our nominal model
averaged over the MACHO field directions
(Fig. \ref{fig:machoave}). About half of the optical 
depths for these directions fall within the $2\sigma$ error bars of
the empirical average optical depth (Fig. \ref{fig:taumacho}), and the
mean optical depth is about $2\sigma$ away from the MACHO mean.  It
is important to remember that although no brown dwarfs are necessary
to match the MACHO data, the number of M stars had to be increased
rather drastically above those obtained from recent star counts both
locally as in the Basu and Rana or Gould {\it et al} mass functions
and in the galactic bulge itself from the Holtzman {\it et al} mass
function.

\noindent 8. For our particular model, the addition of brown dwarfs
to the mass function seems to be a detriment rather than an asset in
matching the MACHO data (Figs. \ref{fig:efmassfns} and
\ref{fig:efmassfns2}).  

\noindent 9. In spite of the success of our nominal model in matching
the MACHO data, uncertainties in all of the galactic parameters
and the meagerness of the published data preclude definitive
microlensing constraints on any of them---including the mass function.
This frustration should be diminished as the microlensing
data set grows.  In fact, additional existing data may already yield
sufficient spatial resolution to constrain the bar shape and
orientation---if the set were analyzed and published.  
 
\begin{center}{\bf Acknowledgements}\end{center}
\acknowledgements
It is pleasure to thank Omer Blaes, Andrew Gould and an unknown
referee for suggestions to improve the manuscript. This work was
supported by the NASA PG\&G program under grant NAGW 2061 and by the
NASA  OSS program under grant NAG5 7177. 
\parindent=0pt
\parskip=10pt
\section{References}
Alcock, C., R.A. Allsman, T.S. Axelrod, D.P. Bennett, K.H. Cook,
K.C. Freeman, K. Griest, J. Guern, M.J. Hehner, S.L. Marshall,
H.-S. Park, S. Perlmutter, B.A. Peterson, M.R. Pratt, P.J. Quinn,
A.W. Rodgers, C.,W. Stubbs, W. Sutherland (1996) The MACHO Project: 45
candidate microlensing events from the first year galactic bulge data,
{\it Astrophys. J.} {\bf 479}, 119-146.

Alvarez, H, J. May, and L. Bronfman (1990) The rotation of the galaxy
within the solar circle, {\it Astrophys. J.} {\bf 348}, 495-502. 

Bahcall, J.N. and R.M. Soneira (1980) The universe at faint
magnitudes. I. Models for the galaxy and the predicted star counts. 
{\it Astrophys. J. Sup.} {\bf 44}, 73-110. 

Basu, S. and N.C. Rana (1992) Multiplicity corrected mass function of
main-sequence stars in the solar neighborhood, {\it Astrophys. J.}
{\bf 393}, 373-384.

Clemens, D.P. (1985) Massachusetts-Stony Brook galactic plane
co-survey: The galactic rotation curve, {\it Astrophys. J.}
{\bf 295}, 422-436.  

Dwek, E. {\it et al} (1995) Morphology, near-infrared luminosity, and
mass of the galactic bulge from COBE DIRBE observations, {\it
Astrophys. J.} {\bf 445}, 716-730.

Gould. A., J. Bahcall, and C. Flynn (1997) M dwarfs from Hubble Space
Telescope star counts .3. The Groth strip, {\it Astrophys. J.} {\bf
482}, 913. 

Han, C. and A. Gould (1996) Statistical determination of the MACHO
mass spectrum, {\it Astrophys. J.} {\bf 467}, 540-545.
 
Han. C. and S. Lee (1998) Inventory of gravitational microlenses
toward the galactic bulge, Submitted to {\it Astrophys. J.}

Holtzman, J.A., A.M. Watson, W.A. Baum, C.J. Grillmair, E.J. Groth,
R.M. Light, R. Lynds, E.J. O'Neil Jr. (1998) The luminosity function
and initial mass function in the galactic bulge, Submitted to {\it
Astrophys. J.}

Kent, S.M., T.M. Dame and G. Fasio (1991) Galactic structure from the
spacelab infrared telescope. 2. Luminosity models of the milky
way, {\it Astrophys. J.} {\bf 378}, 131-138.

Kiraga, M. and B. Paczy\'nski (1994) Gravitational microlensing of the
galactic bulge stars, {\it Astrophys. J.} {\bf 430}, L101-L104.

Kroupa, P. (1995) Unification of the nearby and photometric stellar
luminosity functions, {\it Astrophys J.} {\bf 453}, 358-368.

Mazeh, T., D. Goldberg and D. Latham, (1998) The mass distribution of
extrasolar giant planets and spectroscopic low mass companions, {\it
Astrophys. J. Let.}, In press.

Mera, D., G. Cabrier, and R. Schaeffer (1998) Towards a consistent
model of the Galaxy: I. kinematic properties, starcounts and
microlensing observations,  {\it Astron. Astrophys.} {\bf 330}, 937-952.

Mihalas, D. and J. Binney, (1981) {\it Galactic Astronomy, Structure
and Kinematics}, W.H. Freeman, New York. pp 252, (422). 

Paczy\'nski, B., K.Z. Stanek, A. Udalski, M. Szyma\'nski, J. Kaluzni,
M, Kubiak, M. Mateo, and W. Krzemi\'nski (1994) Are OGLE microlenses
in the galactic bar? {\it Astrophys. J.} {\bf 435} L113-L116. 

Tiede, G.P., J.A. Frogel, and D.M Terndrup (1995) Implications of new
JHK photometry and the deep infrared luminosity function for the
galactic bulge, {\it Astron J.} {\bf 110}, 2788-2812. 

Udalski, A., M. Szyma\'nski, K. Stanek, J. Kaluzni, M. Kubiak, M. Mateo,
W. Krzemi\'nski, B. Paczy\'nski, R. Venkat (1994) The optical
gravitational lensing experiment, {\it Acta Astron.} {\bf 44}, 165

Zhao, H.S. (1996) A steady state dynamical model for the COBE-detected
galactic bar, {\it Mon. Not. R. Astron. Soc.} {\bf 283}, 149-166.

Zhao, H.S., R.M. Rich and D.N. Spergel (1996) A consistent microlensing
 model for the galactic bar, {\it Mon. Not. Roy. Ast. Soc.} {\bf 282},
 175-181.

Zhao, H.S. and T. deZeeuw (1998) The microlensing rate and mass
function vs. dynamics of the galactic bar, {\it
Mon. Not. R. Astron. Soc.} In Press.

Zhao, H.S. D.N. Spergel, R.M. Rich (1995) Microlensing by the galactic
bar, {\it Astrophys. J.} {\bf 440}, L13-L16. 

Zhao, H.S. (1998) Origin of non-unique deprojection of the galactic
bar, {\it Mon. Not. Roy. Astron. Soc.} In Press.  
\end{document}